\documentclass[aps, pre, superscriptaddress, twocolumn,
               floatfix, 10pt,
               longbibliography]{revtex4-1}

\usepackage{hyperref}
\usepackage{natbib}
\usepackage{amsmath}
\usepackage{amssymb}
\usepackage{graphicx}
\usepackage{color}
\usepackage{microtype}

\renewcommand{\vec}[1]{\ensuremath{{\bf #1}}}
\newcommand{\avg}[1]{\ensuremath{\langle #1 \rangle}}
\newcommand{\ket}[1]{\ensuremath{| #1 \rangle}}
\newcommand{\bra}[1]{\ensuremath{\langle #1 |}}
\DeclareMathOperator{\tr}{tr}

\newcommand{\rhodrag}{\ensuremath{\rho_{12}}}

\graphicspath{{img/}}

\begin{document}
\title{Andreev-Bashkin effect in superfluid cold gases mixtures}

\author{Jacopo Nespolo}
\email{e-mail: j.nespolo@lmu.de}

\affiliation{Department of Physics, Arnold Sommerfeld Center for Theoretical
             Physics, Ludwig-Maximilians-Universit\"at M\"unchen,
             Theresienstra{\ss}e 37, D-80333 Munich, Germany, EU}

\author{Grigori E. Astrakharchik}
\affiliation{Departament de F\'isica, Universitat Polit\`ecnica de Catalunya, 
E-08034 Barcelona, Spain, EU}

\author{Alessio Recati}
\email{e-mail: alessio.recati@unitn.it}
\affiliation{Department of Physics, Arnold Sommerfeld Center for Theoretical
             Physics, Ludwig-Maximilians-Universit\"at M\"unchen,
             Theresienstra{\ss}e 37, D-80333 Munich, Germany, EU}
\affiliation{INO-CNR BEC Center and Dipartimento di Fisica, Universit\`a di
Trento, Via Sommarive 14, I-38123 Povo, Italy, EU}

\begin{abstract}
We study a mixture of two superfluids with density-density and current-current 
(Andreev-Bashkin) interspecies interactions.
The Andreev-Bashkin coupling gives rise to a dissipationless drag (or 
entrainment) between the two superfluids.
Within the quantum hydrodynamics approximation, we study the relations between 
speeds of sound, susceptibilities and static structure factors, in a 
generic model in which the density and spin dynamics decouple. 
Due to translational invariance, the density channel does not feel the drag. 
The spin channel, instead, does not satisfy the usual Bijl-Feynman relation, 
since the f-sum rule is not exhausted by the spin phonons. 
The very same effect on one dimensional Bose mixtures and their Luttinger 
liquid description is analysed within perturbation theory.
Using diffusion quantum Monte Carlo simulations of a system of dipolar gases in 
a double layer configuration, we confirm the general results.
Given the recent advances in measuring the counterflow instability, we also 
study the effect of the entrainment on the dynamical stability of a superfluid 
mixture with non-zero relative velocity.

\end{abstract}

\date{\today}

\pacs{}

\maketitle

\section{Introduction}
Mixtures of different kinds of miscible superfluids arise in various areas of
physics, starting from the first experiments on ${}^3$He-${}^4$He mixtures
\cite{PhysRevLett.15.773}, through possible applications to astrophysical
objects \cite{AstrophysJ.282.533, AstrophysJ.836.203}, all the way to the more 
recent developments
in the fields of superconductivity \cite{PhysRevB.89.104508}, cold atoms
\cite{PhysRevLett.89.190404, PhysRevA.77.011603, PhysRevLett.118.055301} and 
exciton-polariton condensates \cite{Science.326.974}.

The statistics of each component of the mixture can be arbitrary, and
Bose-Bose, Bose-Fermi, and Fermi-Fermi mixtures were all successfully realised
experimentally in cold gases. In these experiments, also the chemical nature of
the components can vary: the use of two different elements, of different
isotopes of the same element, and of different internal states of a common
isotope were demonstrated. The ability to reach simultaneous quantum degeneracy
in such a wide variety of atomic species in cold gases experiments allows for
the realisation of very diverse interactions between the two component
superfluids.

One of the most elusive effect 
of coupled superfluids is the existence of a non-zero entrainment between
them. The presence of mutual transport has been pointed out for the first 
time in the 1970s by Andreev and Bashkin \cite{JETP.42.164}, 
correcting some previous work on three-fluid hydrodynamics 
\cite{JETP.5.542,JETP.40.338}. 
The most prominent feature of such an effect, nowadays known as Andreev-Bashkin 
effect (AB), is that the superfluid current $\mathbf{j}_i$ of one component 
will in general depend also on the superfluid velocity $\vec{v}_j$ of the other 
component, or, in other words, that the superfluid density is a non-diagonal 
matrix, $\rho_{ij}$, namely,
\begin{equation}\label{eq:current}
 \vec{j}_i = \rho_{ij} \vec{v}_j,
\end{equation}
with the indices $i,j = \{1, 2\}$ labelling the species and implicit
summation on repeated indices. We shall refer to the off-diagonal element 
$\rhodrag$ of the superfluid density matrix as the \emph{superfluid drag}. 
Phenomenologically, a nonzero $\rhodrag$ carries important implication in the 
dynamics of vortices of a superfluid mixture. Notably, it is predicted that 
the circulation leading to the stable vortex configurations change abruptly as 
$\rhodrag$ is varied, giving rise to stable multiply circulating vortex 
configurations \cite{JETP.60.741}

Despite its introduction was inspired by the problem of $^3$He and $^4$He 
superfluid mixtures, the low miscibility of these two fluids makes this system 
hardly achievable in experiments.
The AB mechanism has instead recently found applications in the domains of 
astrophysics and of 
cold atom systems. In the astrophysical literature, it has been hypothesised 
that the AB effect could be the source of several peculiar behaviours in 
neutron stars cores \cite{AstrophysJ.282.533}, which modern models predict to 
be composed of a mixture of neutrons and protons, both in a superfluid phase
(see \cite{Lattimer536,PhysRevD.70.043001} and reference therein).
Cold atom experiments, on the other hand, thanks to their flexibility and 
tunability, could open the way to a direct measurement of superfluid drag, 
albeit this will still require some careful analysis and mitigation 
of common drawbacks.
For instance, calculations within the Bogoliubov theory for Bose-Einstein 
condensates with typical repulsive interaction---where quantum fluctuations 
are depressed---predict the AB to be very small \cite{LowTempPhys.30.770, 
PhysRevA.72.013616}. 
Quantum fluctuations can be enhanced by increasing the interactions, but this 
would also intensify three-body losses, which could be in turn suppressed by 
confining the system in low dimensional geometries or introducing an optical 
lattice, as 
studied, e.g., in \cite{PhysRevA.79.063610, PhysRevA.86.033627}.
However, optical lattices break translational invariance, thus strongly
reducing the superfluid density even a $T=0$. We recall
that, in continuous space (and with time-reversal symmetry), the superfluid
density approaches the total density as $T$ is lowered to zero (see, e.g.,
\cite{Leggett1998}).

Aside from making the AB mechanism efficient, a very important question is 
how to measure experimentally its strength. The dynamical protocols typically 
proposed require the ability to initialise a superfluid current in one 
component and then observe the onset of dissipationless transport in the other 
one, initially at rest.
For best results, these kind of measurements would likely require a ring 
geometry and can be of difficult interpretation, since a number 
of decay processes are present \cite{ZoranCurrent, AlessioCurrent}.

In the present work, we address some of the above mentioned issues. In
particular, we derive some relations between the superfluid drag and other
measurable quantities, such as the susceptibilities of the system and the
speeds of sound.
For systems with $\mathbb{Z}_2$ symmetry between the two species, in which spin 
and density channels decouple, the density channel follows the usual relations,
whereas we show how the AB breaks the usual Bijl-Feynman relation for the spin 
channel. Our findings open the way to measuring the superfluid drag 
experimentally using standard static and dynamic observables.

To provide support to our theoretical predictions, we study quantitatively 
a specific model which can show large entrainment, i.e., a dipolar Bose gas 
trapped in a bilayer configuration. 
Using the diffusion quantum Monte Carlo method, we extract the dispersion 
relations, the 
susceptibilities, the structure factors and the superfluid densities. 
We show that they satisfy, in a proper regime, the expression derived in the 
general theory.
In particular, it is shown that the standard expression relating the
square of the spin speed of sound to the inverse of the susceptibility is
inapplicable and it should be corrected by a factor proportional to
the superfluid drag.
Another possible quantity which could reveal the presence of a superfluid drag 
is the shift in the position of the dynamical instability.
We report a general stability analysis of the mixture, 
and derive a simple analytical expression for the onset of the dynamical 
instability to linear order in the drag.

As mentioned above, the AB mechanism could play a more prominent role in low 
dimensionality. We discuss the modifications to Luttinger liquid 
theory necessary to describe coupled one dimensional superfluids.
We find that, in analogy with the general description, the spin Luttinger 
parameters, as derived by means of perturbative or ab-initio calculations,
receive a correction from the superfluid drag, which could become particularly 
relevant in the strongly interacting (Tonks-Girardeau) regime.

The paper is organised as follows.
In Sec.~\ref{sec:ab_intro} we recall the main aspects of the AB effect, which 
are then analysed within a minimal quantum hydrodynamic toy model in 
Sec.~\ref{sec:quantum_hydro}.
The relations among experimentally relevant observables are derived. The 
Luttinger liquid theory for one dimensional coupled superfluid is corrected for 
the presence of AB in the same section.
Numerical evidence in support of our theoretical findings are reported in 
Sec.~\ref{sec:numerics}, where we analyse the presence and magnitude of the 
superfluid drag in a bilayer system of dipolar bosons.
In Sec.~\ref{sec:dynamic_stability} we study the dynamical instability of the 
mixture with respect to the relative velocity between the two fluids.
Conclusions and future perspectives are drawn in Sec.~\ref{sec:conclusions}. 
For the sake of completeness, the derivations of some relations used in
the main text are postponed to the appendices without affecting the
comprehension of the main results.

\section{Andreev-Bashkin effect}\label{sec:ab_intro}
Microscopically, the current drag originates from the interactions between two 
superfluids, leading to the formation of quasi-particles with nonzero content 
of either of the two species. It is then easy to understand that the transport 
properties of the two components are not independent: the flow of one 
component must be accompanied by mass transport of the other component 
\cite{JETP.42.164}.

Some important relations concerning the superfluid densities in
Eq.~\eqref{eq:current} can be easily obtained by considering the kinetic
energy contribution in the expansion of the ground state energy in terms of
the superfluid velocities \cite{PhysRevLett.95.090403}. Due to Galilean
invariance, if $\phi_i$ is the phase of the superfluid order parameter for
component $i$, of mass $m_i$, its velocity is given by $\vec{v}_i= (\hbar/m_i)
\nabla \phi_i$ and the energy due to the superfluid velocities can be written
as
\begin{equation} \label{eq:energy_Y}
 \delta E = \int d^D x\,  \sum_{ij}
\frac{\hbar^2}{2m_i m_j}\rho_{ij} \nabla \phi_i \cdot
\nabla \phi_j.
\end{equation}
By performing a Galilean boost with velocity $\vec{V}$, the phases are shifted 
to
$\phi'_i = \phi_i - (m_i / \hbar)\vec{V} \cdot \vec{r}$, and the
energy change, to first order in $\vec{V}$, is $\delta E' = \delta E - \int
\vec{P}\cdot\vec{V}\, d^Dx$, with $\vec{P}/\hbar = \sum_i
(\rho_i + \rhodrag) / m_i \nabla \phi_i $ the momentum density. Since
on the other hand one must have $\vec{P}=n_1 \nabla \phi_1 + n_2 \nabla
\phi_2$, with $n_{1,2}$ the number densities, the superfluid densities must
satisfy
\begin{equation}\label{eq:density_condition}
m_i n_i = \rho_i + \rhodrag.
\end{equation}
Introducing the effective masses $m_{1,2}^*$ through $\rho_{ii}
\equiv n^{}_i m_i^2/ m_i^*$, we obtain
\begin{equation}\label{eq:drag_mstar}
 \rhodrag = n_i m_i \left( 1 - \frac{m_i }{ m^*_i} \right),
\end{equation}
which provides the relation between the superfluid drag and the effective
masses and a constraint for the effective mass ratio.

\section{Quantum hydrodynamic model} \label{sec:quantum_hydro}

In general, the presence of effective masses changes the relation between 
static and dynamical properties of the system, and the possibility for some 
excitation modes to exhaust the sum rules. Let us consider a
superfluid mixture with an energy density $e(n_1, n_2)$. Expanding the energy
around its ground state value to second order in the density fluctuations
$\Pi_i(x)$ and adding it to Eq.~\eqref{eq:energy_Y} we obtain a hydrodynamic
Hamiltonian for two miscible superfluids,
\begin{equation}\label{eq:Hhydro}
 H=\frac{1}{2}\sum_{ij}\int\left(
 \rho_{ij} \frac{\hbar \nabla \phi_i}{m_i} \cdot
             \frac{\hbar \nabla \phi_j}{m_j}
+\alpha_{ij}\Pi_i\Pi_j\right)\, d^D x,
\end{equation}
where the matrix $\alpha_{ij}=\partial^2e/\partial n_i\partial n_j$ contains 
the information on inter- and intra-species interactions of the two fluids. 
Hamiltonian~\eqref{eq:Hhydro} by requiring that the fields 
$\phi_i$ and $\Pi_j$ satisfy canonical commutation relations for bosons, i.e., 
$[\phi_i(x), \Pi_j(y)] = i\hbar \delta_{ij}\delta(x-y)$.

For the sake of clarity, we take the two superfluids to be equal:
$\rho_{ii}=\rho$, $\alpha_{ii}=\alpha$, $m_i = m$ and $n_i = n/2$, with $n 
\equiv N/V$ the total number density of the system (see 
Appendix~\ref{app:nonZ2} 
for the non-symmetric case). Due to the 
assumed $\mathbb{Z}_2$ symmetry, the dynamics of this model decouples
if we rewrite it in terms of the new fields
\begin{equation}
 \phi_{d(s)} = (\phi_1 \pm \phi_2)/\sqrt{2}, \quad
 \Pi_{d(s)} = (\Pi_1 \pm \Pi_2)/\sqrt{2}.
\end{equation}
The fields $\Pi_d$ and $\phi_d$ represent the fluctuations in total density 
and global phase, respectively. In a similar fashion, $\Pi_s$ and $\phi_s$ 
encode the fluctuations of the difference in density of the two species 
(magnetisation) and their relative phase (spin wave), respectively. We use the 
labels $d(s)$ to indicate the density (spin) channel of the system's 
excitations. The new fields inherit the canonical commutation relations and 
act as two independent hydrodynamic modes, obeying the Hamiltonian
\begin{equation}\label{eq:Hhydro_normal_modes}
 H = \frac{1}{2} \sum_{i=d,s} \int
 \left[\rho_{i} \left(\frac{\hbar\nabla\phi_i}{m}\right)^2 +
       \alpha_{i} \Pi_i^2 \right] \, d^Dx,
\end{equation}
where $\rho_{d(s)}=\rho \pm \rhodrag$ and $\alpha_{d(s)}=\alpha\pm 
\alpha_{12}$. 
The Hamiltonian is now diagonal in the two channels, and the dispersion 
relations for the two modes are linear in the momentum $k$, of the form
\begin{align}\label{eq:hydro_dispersion}
 (\hbar \omega)^2 = \frac{\alpha_i \rho_i}{m^2} (\hbar k)^2, \qquad (i=d,s).
\end{align}
The quantity $\alpha_i \rho_i / m^2$ can be identified with the speed of sound
of each mode. For the density mode, we have
\begin{equation}\label{eq:cd}
 c_d^2 = \frac{(\rho + \rhodrag)}{m^2} (\alpha + \alpha_{12})
          = \frac{n}{2m}(\alpha + \alpha_{12}),
\end{equation}
where in the last equality we used that, at $T=0$, the total superfluid 
density is
equal to the total mass density of the system. We note in particular that
$c_d$ is independent of the superfluid drag. On the other hand, the spin speed
of sound is
\begin{equation}\label{eq:cs}
 c_s^2 = \frac{(\rho - \rhodrag)}{m^2} (\alpha - \alpha_{12})
       = \frac{n}{2m}\left(\frac{2m}{m^*} - 1 \right)(\alpha - \alpha_{12}),
\end{equation}
which explicitly depends on $\rhodrag$.

From the Hamiltonian \eqref{eq:Hhydro_normal_modes}, the static response to 
density and to spin probes are simply given by $\alpha_{d(s)}$, which 
can be identified with the inverse compressibility $\kappa_d^{-1}$ and inverse 
magnetic 
susceptibility $\chi_s^{-1}$, respectively. We thus obtain the relations
\begin{align}
 c_d^2 &= \frac{n}{2m\kappa_d}, \label{eq:cd_kappa} \\
 c_s^2 &= \frac{\rho-\rhodrag}{m^2\chi_s}
       = \frac{n}{2m\chi_s} \left(\frac{2m}{m^*} - 1 \right).
       \label{eq:cs_chi}
\end{align}
These relations suggest that, by independently measuring $c_s$ and 
$\chi_s$, it is possible to obtain the strength of the mass renormalisation, 
i.e., the magnitude of the superfluid drag.
Note that $c_s$ is expected to vanish for $m^* = 2m$, which imposes a bound 
$m^* \leq 2m$. From Eq.~\eqref{eq:drag_mstar}, this bound translates into 
$\rhodrag \leq mn / 4$, thus anticipating result~\eqref{eq:drag_bound}, of 
which we will provide an additional derivation below.

The previous analysis has important consequences with respect to Bijl-Feynman 
relations (f-sum rule) linking the dispersion relations to the static 
structure factors (see, e.g., \cite{GiulianiVignale}). From the above 
discussion, it turns out that the f-sum rule for the density channel is 
exhausted by the phonon mode, while for the spin mode this is not the case, 
leading to the effective mass correction in the determination of the 
dispersion relation. In particular, the zero temperature spin structure factor 
at low momenta reads
\begin{equation}\label{eq:Ssk_vs_chi}
S_s(k) \stackrel{k\to0}{=} \frac{k}{2 m c_s} \frac{(\rho - \rhodrag)}{m}
       =\frac{k}{2m} \sqrt{(\rho - \rhodrag) \chi_s},
\end{equation}
which does not satisfy the Bijl-Feynman relation. Notice that the linear term
in $k$ of the $S_s(k)$ can vanish, either because of a vanishing susceptibility
or because of a saturated drag, i.e., $\rhodrag = \rho$. The former (latter)
case corresponds to a vanishing (diverging) spin speed of sound.
The fact that the drag and the interspecies interaction act independently 
on the spin speed of sound [cf.\ Eq.~\eqref{eq:cs}] is general, and applies 
beyond the $\mathbb{Z}_2$ symmetry we assumed in this section. In 
particular, the standard condition for the onset of phase separation (i.e., 
$\chi_s\rightarrow \infty$ for $\alpha=\alpha_{12}$) still holds 
(cf.\ Appendix~\ref{app:nonZ2}). 

On the other hand, due to translational invariance, the density structure 
factor satisfy the Bijl-Feynman relation and it reads
\begin{equation}\label{eq:Sdk_vs_kappa}
S_d(k) \stackrel{k\to0}{=} \frac{n k}{4 m c_d} 
       =\frac{k}{2m} \sqrt{2m n \kappa_d}.
\end{equation} 
 
Let us conclude this section by briefly mentioning the effect of the AB 
physics on the specific heat of the mixture.
At low but finite temperature, we may expect that thermal fluctuations do not 
change the low energy spectrum significantly. 
Then the low temperature dispersion relations are still linear, 
of the form, $\epsilon_i(k) = c_i k$, $(i=d,s)$, and we assume, within the 
hydrodynamic picture, that the highest momentum that can be thermally excited 
is $k_{T,i} = k_BT/c_i$. In the low temperature limit and $D$ spatial 
dimensions, these assumptions lead to the specific heat
\begin{equation}
 C_v \propto T^{D}\left(\frac{1}{c_d^D}+\frac{1}{c_s^D}\right),
\end{equation}
which carries a dependence on $\rhodrag$ through the sound velocity in the 
spin channel. A large superfluid drag will therefore lead to a strong increase 
of the specific heat. 

\subsection{One dimensional systems and Luttinger Liquid}
Since the superfluid drag is due to quantum fluctuations, one can think about 
increasing them by increasing the interactions, i.e., quantum depletion and 
mutual dressing.
This can be easily seen in the weakly interacting regime, where analytical 
expressions for the superfluid drag have been nicely obtained within a 
Bogoliubov approach by Fil and Shevchenko \cite{PhysRevA.72.013616, 
LowTempPhys.30.770}.
In a three-dimensional system, three-body losses strongly limit the possible 
increase of the interaction strengths.
On the other hand, in one dimension, it is possible to reach strong quantum 
regimes, including the so-called Tonks-Girardeau regime.
The low-energy excitations of one-dimensional gases are described in terms of 
Luttinger liquids \cite{Giamarchi}.
For the sake of simplicity, we consider two equal Luttinger liquids coupled 
together, with speed of sound $c_0$ and Luttinger parameter $K_0\ge 1$.
By introducing both the density-density and the current-current  couplings as a 
perturbation, we can write
\begin{eqnarray}
H_{LL}&=&\sum_{i=1,2} \frac{c_0}{2}
\int [K_0 (\partial_x \phi_i)^2 + \frac{1}{K_0} \Pi_i^2] \nonumber \\
 &+& \int [\rhodrag \frac{\partial_x \phi_1 \partial_x \phi_2}{m^2} 
      + g_{12} \Pi_1 \Pi_2].
\label{eq:H_luttinger}
\end{eqnarray}
As before, we can easily diagonalise the Hamiltonian \eqref{eq:H_luttinger} 
by introducing the fields for the in-phase and out-of-phase fluctuations.
We obtain a standard expression for coupled Luttinger liquids
\begin{equation}
H_{LL}=\sum_{i=d,s} \frac{c_i}{2}
\int [K_i (\partial_x \phi_i)^2 + \frac{1}{K_i} \Pi_i^2],
\end{equation}
where the density parameters read
\begin{align}
2mc_{d}^2 &= n (c_0/K_0+g_{12}), \\
2mK_{d}^2 &= n/(c_0/K_0+g_{12}),
\end{align}
and for the spin sector we get
\begin{align}
\label{spinLL}
2mc_{s}^2 &= (n-4\rhodrag/m)(c_0/K_0-g_{12}), \\
2mK_{s}^2 &= (n-4\rhodrag/m)/(c_0/K_0-g_{12}).
\end{align}
In the above expressions, $n$ is the total density of the system,
and we have used the fact that, for a translationally invariant 
system, $c_d K_d = n / 2m$ (see also \cite{condmat.9807366}),
which implies that $c_0 K_0 + \rhodrag / m^2 = n / m$.
Therefore, $c_d$ and $K_d$ do not depend on the off-diagonal superfluid density
and for the compressibility we have $\kappa = K_d / c_d = n / (2m c_d^2)$.
On the other hand, the spin channel parameters acquire a dependence on 
$\rhodrag$, as seen before in the general case. In fact, the susceptibility 
reads $\chi_s = K_s / c_s = (n - 4 \rhodrag) / (2 m c_s^2)$, to be compared 
with Eq.~\eqref{eq:cs_chi}.
The correction due to AB in Eq.~\eqref{spinLL}, in the strongly 
interacting limit can therefore deeply modify the standard 
perturbative analysis \cite{NewJPhys.10.045025,PhysRevA.77.013607} and the RG 
flow for coupled Bose Luttinger liquids.
Note, once again, that the previous equations imply a bound on the value of the 
superfluid drag, $\rhodrag \leq nm / 4$, which coincides with the one coming 
from Eq.~\eqref{eq:cs_chi} in the previous section.
Recent Monte-Carlo simulations on one dimensional Bose gases  confirm our
results \cite{parisi_giorgini_private}.

\section{Magnitude of the drag and numerical evidence} \label{sec:numerics}
The information on the superfluid drag can be extracted from quantum
Monte-Carlo (QMC) simulations based on the path integral formalism. In this
formalism, in fact, the superfluid density can be related to the statistics of
winding numbers of particles' paths around the simulation domain
\cite{PhysRevB.36.8343}. By extending this result to two species in the same
simulation box (see Appendix~\ref{sec:app_path_integral} for details), we
obtain the relation
\begin{align} \label{eq:winding_drag}
 \rho_T & = \frac{L^{2-D}}{\beta D}
      \left[m_1^2 \avg{W_1^2} + m_2^2 \avg{W_2^2} + 2 m_1 m_2 \avg{W_1 W_2}
      \right] \nonumber \\
  & = \rho_{1} + \rho_{2} + 2 \rhodrag,
\end{align}
linking the total superfluid density $\rho_T$ to the winding numbers $W_{1,2}$
of the two species. Here we are considering a simulation volume $L^D$ at
inverse temperature $\beta = 1/T$. For zero temperature results, $T$ is taken
smaller than all the other energy scales of the system and the results
are checked \emph{a posteriori} for convergence.

From Eq.~\eqref{eq:winding_drag}, the superfluid drag can be interpreted as the
covariance between the superfluid densities of the two components. Then, thanks
to Cauchy-Schwarz inequality, $\rhodrag^2 \leq \rho_1 \rho_2$, and assuming the
symmetric case, in which $\rho_1 = \rho_2$, we obtain an upper bound on the 
magnitude of the drag, 
\begin{equation}\label{eq:drag_bound}
 \rhodrag \leq \frac{\rho_T}{4}, \qquad (\rho_1 = \rho_2)
\end{equation}
which also bounds the effective mass to $m^* \leq
2m$. As already noted above, the condition of saturation of this bound 
corresponds to a vanishing speed of sound in the spin channel [cf.\ Eqs.\
\eqref{eq:cs}-\eqref{eq:cs_chi}]. We shall also see below that the saturation
of this bound is a limiting case in the dynamic stability of the mixture [see 
Eq.~\eqref{eq:vcrit_z2}].

\subsection{Quantum Monte Carlo results for bilayer dipolar Gases}

Lattice simulations already showed evidence of superfluid drag effects
\cite{PhysRevA.86.033627, PhysRevLett.95.090403}. It was shown that the drag
depends on the lattice geometry, increases with the increase of interspecies
interactions and attains its maximum for non-equal masses of the two particle
species. The presence of the lattice explicitly breaks the translational
invariance, thus deeply modifying the mechanism leading to a dissipationless
drag. In particular, for incommensurate fillings, the drag between the two
fluids is essentially mediated by the presence of vacancies
\cite{PhysRevLett.95.090403}.

The magnitude of the superfluid drag, normalised by the total superfluid
density, spans the whole range allowed by bound \eqref{eq:drag_bound}.
However, one must point out that the presence of the lattice causes the
depletion of the total superfluid density; in particular, on a lattice, it is
no longer true that the total superfluid density coincides with the total
particle density at zero temperature \cite{Leggett1998}. In this context, it is
noteworthy to mention the analytical results of Ref.~\cite{PhysRevA.79.063610},
which compute the superfluid drag starting from the physical parameters of the
lattice in a weak coupling approximation. The authors report a superfluid drag
$\rhodrag / nm$, normalised to the total mass density, of the order of
$10^{-5}$--$10^{-4}$ for weak to moderate intercomponent scattering amplitude.
Quantitatively similar QMC results are reported in \cite{PhysRevA.86.033627}.
It is important to keep in mind that these low values are primarily due to the
small total superfluid density on the lattice.

\begin{figure}
 \includegraphics{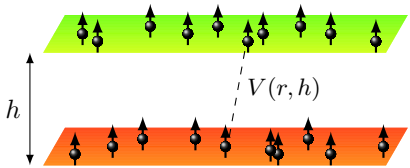}
 \caption{Schematic representation of the bilayer dipolar bosonic model. 
Particles obeying Bose-Einstein statistics are confined in two parallel layers. 
The dipoles are pinned perpendicularly to the layers' planes and parallel to 
each other, yielding entirely repulsive on-plane interactions and 
partially attractive out-of-plane interactions.}
\label{fig:BDB}
\end{figure}

\begin{figure}[tb!]
\includegraphics{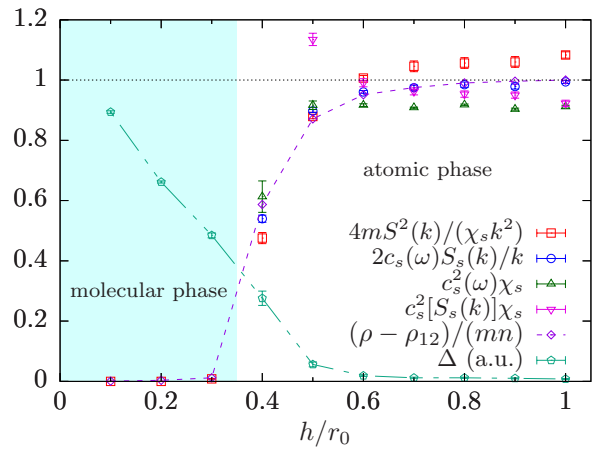}
\caption{
The quantity $(\rho-\rhodrag)$, extracted from diffusion QMC data using 
different estimators. The direct winding number estimator (diamonds and dashed 
line to guide the eye) is compared with indirect estimators. The latter make 
use of the relations derived in the quantum hydrodynamic model of Sec.\ 
\ref{sec:quantum_hydro} (cf.\ also Eq.~\eqref{eq:many_rho12}). The data sets are 
in satisfactory agreement with one another, with the exception of the estimator
$c_s^2 [S_s(k)] \chi_s$, which tends to diverge as the molecular phase is 
approached (shaded region). The origin of the errorbars and of remaining 
finite-size effects are discussed in the main text.}
\label{fig:dqmc}
\end{figure}

In the following, we focus on a system of dipolar Bose gases confined in a 
bilayer geometry in continuous space, with the dipole orientation pinned 
perpendicular to the planes, as sketched in Fig.~\ref{fig:BDB}.
This system is similar to the one studied in Ref.~\cite{LowTempPhys.25.493}, 
which pointed out the presence of entrainment between the superfluid currents 
of two charged superfluids in a bilayer configuration.
The relative strength of interspecies interactions as compared to intraspecies 
ones can be tuned by changing the distance between the two layers.
As it will be shown later in Fig.~\ref{fig:dqmc}, for an extended range of 
this control parameter, the superfluid drag can reach very large values.
Dipolar particles in a bilayer configuration are particularly advantageous 
under a variety of aspects.
Confining the molecules in a two-dimensional geometry and imposing a repulsive 
dipolar interaction strongly reduces the detrimental two-body chemical 
reactions \cite{Ni2010}.
At the same time it allows to exploit the anisotropy of the 
dipolar interaction, which is partially attractive between particles on 
different layers. Introducing the distance $h$ between the two layers, the 
interaction between two particle of mass $m$ and dipole moment $d$ on different 
layers can be written as
\begin{equation}\label{eq:dipolar_potential}
V(r,h)=d^2\frac{r^2-h^2}{(r^2+h^2)^{5/2}},
\end{equation}
where $r$ is the relative distance in the plane of motion. 
For dipolar gases, it is very useful to introduce the characteristic 
length $r_0=md^2/\hbar^2$. The various regimes of the system
are characterised by the interlayer parameter $h/r_0$ 
and the in-layer parameter $n_i r_0^2$, with $n_i$ the single layer density.
Static and dynamic properties of this system were recently investigated in
\cite{PhysRevA.90.043623, PhysRevA.94.063630}. 
In particular, it has been found that a transition from two 
coupled superfluid (atomic phase) to a pair superfluid (molecular phase) takes 
place when the attractive interaction is strong enough.
We will show that, by approaching the transition point while remaining in the 
atomic phase, the drag superfluidity becomes prominent.

To recover the description of Eq.~\eqref{eq:Hhydro}, we point out that 
miscibility is here to be intended with respect to the position of the 
particles projected in the direction orthogonal to the layers' planes. Dipolar 
Bose gases in a double layer configuration do not show any phase separation 
\cite{PhysRevA.90.043623}. This is intuitive, since any potential with 
$V(q)|_{q=0}\leq 0$ (in momentum space) admits a bound state in $2D$. (The 
interlayer potential \ref{eq:dipolar_potential} has the peculiarity to have 
$V(q)|_{q=0}=0$, which makes the bound and scattering states of the system at 
weak coupling rather peculiar \cite{PhysRevA.82.044701}). The dipolar bilayer 
Bose gas is moreover the first example of a two component Bose gas which can 
form pairs without collapsing (i.e. forming clusters) \cite{PhysRevA.90.043623} 
as it occurs, e.g., in mixtures with contact interaction only.

Besides serving as a testbed for the numerical study of the superfluid drag in 
a homogeneous geometry and being a new system showing the AB physics, the 
dipolar bilayer configuration can represent one of the best-case scenarios for 
the experimental observation of the presence of superfluid drag.
Recent experiments using dipolar molecules consisting of two atoms of 
Erbium-168 demonstrated the availability of condensates with large magnetic 
moments, up to $r_0 \approx 1600\, {\rm a}_0$, with ${\rm a}_0$ the Bohr radius 
\cite{PhysRevLett.115.203201}. This value is still almost one order of 
magnitude smaller with respect to the typical wavelength of 
the lasers used to confine the Er$_2$ molecules in arrays of 2D layers.
Experiments on stable polar Na-K molecules \cite{MartinNaK}, which sport much
larger $r_0$, may help in overcoming this problem.
The recent proposal of sub-wavelength confinement 
\cite{PhysRevLett.115.140401} may further stretch the experimentally accessible 
range of values of $h/r_0$, albeit it is not of easy implementation for dipolar 
molecules.
Therefore, experimental realisation of a bilayer system of strongly 
interacting dipolar superfluids is reasonably within reach of current or 
near-future technology.

We study the system by means of diffusion QMC, which allows us to extract 
both the thermodynamics and the low energy spectrum of the system.
Diffusion QMC is based on solving the Schr\"odinger equation in imaginary 
time, thus projecting out the ground state of the system (for a general 
introduction on the method see, e.g., \cite{PhysRevB.49.8920}). The 
contributions of the excited states are exponentially suppressed and the 
ground-state energy is recovered in the limit of long propagation time. The 
simulations are performed for $60$ particles with the same parameters as in 
Refs.~\cite{PhysRevA.90.043623, PhysRevA.94.063630}. For some quantities, this 
number of particles is sufficiently large to be close to the thermodynamic 
limit; for some others, residual finite-size corrections must be taken into 
account, as it will be explained in more details later.

In this framework, a number of observables of interest can be obtained in a 
straightforward way. The value of the gap $\Delta$ and of the spin 
susceptibility $\chi_s$ are obtained from the dependence of the ground-state 
energy on the polarisation $P = (N_1-N_2) / N$. The latter is tuned by moving 
particles from layer $1$ to layer $2$ while keeping the total number of 
particles $N = N_1 +N_2$ constant. In the limit of small polarisation $P$, the 
energy can be expanded as
\begin{equation} \label{eq:susceptibility}
E(P) = E(0) + N \Delta \cdot P + N \frac{n}{2 \chi_s} \cdot P^2.
\end{equation}
In the gapless phase ($\Delta=0$) the dependence on the polarisation is
quadratic, while in the gapped phase it is linear.
Similarly, the 
compressibility $\kappa_d$ at $T=0$ can be obtained from the volume dependence
of the energy for an unpolarised gas,
\begin{equation} \label{eq:compressibility}
\kappa_d^{-1} = -\mathcal{V} \left(
     \frac{\partial^2 E}{\partial \mathcal{V}^2}
\right),
\end{equation}
where $\mathcal{V}$ is the $D$ dimensional volume of the system ($\mathcal{V} 
= L^2$ in the $2D$ geometry at hand).

The study of structure factors provides a way of accessing the dynamic 
properties of the system. We use the technique of pure estimators 
\cite{PhysRevA.10.303, PhysRevB.52.3654} to compute the intermediate scattering 
function
\begin{equation}
 S_{\alpha\beta}({\bf k},\tau) = \frac{1}{N}
     \left\langle
          \rho_\alpha (\vec{k}, \tau) \rho_\beta(-\vec{k}, 0) 
     \right\rangle,
\end{equation}
with $\rho_\alpha(\vec{k}, \tau) = \sum_j^{N_\alpha} \exp\lbrace -i\vec{k} 
\cdot \vec{r}_{j\alpha}(\tau) \rbrace$ and $\vec{r}_{j \alpha}$ the position 
of particle $j$ in layer $\alpha$.
The intermediate scattering function provides information on 
the correlations in imaginary time $\tau$ and is the main ingredient to 
compute 
the static structure factor $S_{\alpha \beta}(\vec{k}) \equiv S_{\alpha 
\beta}(\vec{k}, 0)$. We consider a balanced system with $N_A = N_B$ and study 
the symmetric and antisymmetric structure factors, 
\begin{equation}
 S_{d(s)}(k) = S_{11}(k) \pm S_{12}(k),
\end{equation}
corresponding to the density and spin channels of the discussion above, 
respectively.
The compressibility and the spin susceptibility can be compared to the 
respective static structure factors in the low momentum limit, in order to 
verify the sum rules. The structure factors further provide information on the 
excitation spectra: their long imaginary time asymptotic behaviour can be 
fitted to an exponential decay of the form
\begin{equation}\label{eq:omega_decay}
 S_{d(s)}(\vec{k}, \tau) \sim Z e^{ -\omega_{d(s)}(k) \tau}.\qquad
(\tau\to\infty)
\end{equation}
When phononic excitations are present, $\omega_{d(s)}(k)$ is linear for small
momenta, with the slope directly related to the speeds of sound of the density
and spin channels, respectively, through $\hbar \omega_{d(s)}(k) \simeq
c_{d(s)} k$.

\begin{figure}[tb!]
\includegraphics[width=8.5cm]{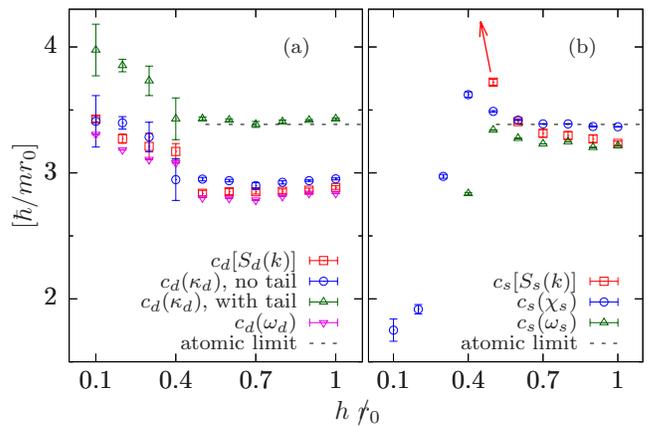}
\caption{Speeds of sound as a function of interlayer spacing $h$ for $nr_0^2 = 
1$, as extracted from different observables.
(a) Speed of sound of the density mode, $c_d$.
(b) Speed of sound of the spin mode, $c_s$.
The Feynman method makes use of the static structure factor $S(k) = \hbar k / 
(2mc)$, computed at the smallest $k$ compatible with periodic boundary 
conditions.
The speeds $c_{\alpha}[S_{\alpha}(k)]$, with $\alpha = \lbrace d,s \rbrace$ 
the channel index, are computed with this method.
The data show that the f-sum rule is exhausted by the phonon mode in the 
density channel, whereas this does not hold in the spin channel, the arrow 
indicating the divergence of $c_s[S_s(k)]$.
The speeds $c_{\alpha}(\omega_\alpha)$, computed from the excitation spectrum,
assume a linear phononic dispersion relation $\omega(k) = ck$ with 
$\omega_\alpha(k)$ obtained from Eq.~\eqref{eq:omega_decay}.
The speeds $c_d(\kappa_d)$ and $c_s(\chi_s)$ are computed from $mc_d^2 = n 
\kappa_d^{-1}$ and $mc_s^2 = n \chi_s^{-1}$, with $\kappa_d$ and $\chi_s$ 
obtained from Eqs.~\eqref{eq:compressibility} and \eqref{eq:susceptibility}, 
respectively.
The speed of sound in the atomic limit coincides in the two channels. It is 
obtained from standard thermodynamic relations using the equation of state 
$E(nr_0^2)$ (taken from Ref.~\cite{PhysRevA.75.063630}) of a single layer 
system with half the density of the bilayer system.}
\label{fig:dqmc:c}
\end{figure}

It is instructive to show that, in a gapless system without the drag, exactly 
the same information on the speeds of sound can be recovered from the static 
structure factors $S_{d,s}(k)$, the low-momentum excitation spectra 
$\omega_{d,s}(k)$, the 
compressibility $\kappa_d$ and the susceptibility $\chi_s$. The speeds of sound 
obtained from the different methods are shown in Fig.~\ref{fig:dqmc:c} for the 
density~(a) and spin~(b) modes. The density mode is gapless for any value of 
the interlayer separation $h$. The speed of sound of this channel, as obtained 
from structural, energetic and thermodynamic quantities, always yields 
compatible values throughout the explored range of $h$. Finite-size effects 
reduce the speed of sound, which for large $h$ (decoupled layers) appears to 
lie below its asymptotic value. The latter is obtained from the equation of 
state of a model with a single species at half the density (so called 
``atomic'' limit). In the computations, the dipolar 
interaction potential was truncated at a distance equal to half the size of 
the simulation box. By adding the missing ``tail'' correction to the 
compressibility, it is possible to recover correct atomic limit asymptotics, 
as shown in the figure.
The situation is quite different for the spin channel, where the gap opens for 
$h/r_0 \lesssim 0.35$ and different methods cannot be consistent in that 
parameter range. For large values of the interlayer separation, $h/r_0 \gtrsim 
0.6$, we recover once again the atomic limit and different quantities are 
consistent with one another. Manifestly, it is not the case for parameter 
range $0.35 \lesssim h / r_0 \lesssim 0.6$, which still corresponds to a 
gapless phase but is in the vicinity of the transition point. In this region 
there is no consistency between the speeds of sound obtained with different 
methods and, importantly, the f-sum rule is not satisfied. The reason for this 
is appearance of the superfluid drag which we analyse in more details below. It 
is interesting to note that the finite-size effects are more pronounced in the 
density mode compared to the spin mode. One way to understand this is that the 
spin mode probes the response to the polarisation, which does not change the 
system volume, while the density (compression) mode is the response to a change 
in volume. The tail correction, being sensitive to the change of the volume, is 
able to account for the finite-size discrepancy.

Finally, in order to directly probe the superfluidity properties of the 
system, we introduce the winding number related to species $\alpha$,
\begin{equation}
 \vec{W}_{\alpha}(\tau) = \sum_{i(\alpha) = 1}^{N/2} \int_0^\tau d\tau'
\frac{d\vec{r}_{i(\alpha)}(\tau')}{d\tau'},
\end{equation}
where $i(\alpha)$ indexes particles belonging to species $\alpha$ only.
Taking the limit of long propagation time, the statistics of the winding
numbers are related to the superfluid densities. In particular, we evaluate the
symmetric and antisymmetric combinations
\begin{equation} \label{eq:qmc_winding}
\rho_1 + \rho_2 \pm 2\rhodrag = \lim_{\tau \to \infty}
     \frac{\langle [\vec{W}_1(\tau) \pm \vec{W}_2(\tau)]^2 \rangle}{2 N \tau}.
\end{equation}
According to Eq.~\eqref{eq:winding_drag}, when the plus sign is considered, the
quantity above is an estimator of the total superfluid density of the system,
$\rho_T$. In our zero temperature simulation, this quantity is always
compatible with the total mass density, as it should be in continuous space, 
and in contrast
with the low total superfluidity observed in lattice simulations. When the
minus sign is considered, on the other hand, we directly probe the magnitude of
the superfluid drag. This observable asymptotically attains the value $\rho_T$
in the non interacting ($h \to \infty$) limit. In the symmetric mixture case,
borrowing the same notations of Sec.~\ref{sec:quantum_hydro}, this quantity
reduces to $(\rho - \rhodrag) / nm$, and is reported in Fig.~\ref{fig:dqmc}.
For low interactions (large $h$), it is compatible with unity and drops to zero
as interactions are ramped up. This corresponds to $\rhodrag = \rho_T / 4$, 
i.e., with bound \eqref{eq:drag_bound}. It must be noted that, for $h < h_c 
\approx 0.35 r_0$\cite{PhysRevA.90.043623}, the system enters the molecular 
phase, so that not all the drop in $\rho - \rhodrag$ can be ascribed to an 
increase of the drag, but one must also keep into account the emergence of the 
molecular condensate. Indeed, the description we put forward holds only as long 
as the system is still in the atomic phase. Interestingly, the saturation of 
the bound \eqref{eq:drag_bound} (or, alternatively $m^*\rightarrow 2m$) appears 
to coincide with the transition to the molecular phase in which bound state 
physics start dominating.

By independently measuring $S_s(k)$, $\chi_s$ and $\omega_s$ (hence $c_s$), we
can show that the usual Bijl-Feynman approximation is not applicable to
systems in the presence of the superfluid drag.  To this end, we use Eqs.\
\eqref{eq:cs_chi}-\eqref{eq:Ssk_vs_chi} to express $(\rho-\rhodrag)$ in terms
of the observables listed above, and then compare it with the direct winding
number measurement.  Figure \ref{fig:dqmc} reports the data corresponding to
the three independent expressions
\begin{equation} \label{eq:many_rho12}
\frac{\rho-\rhodrag}{nm}
 = \frac{4mS^2_s(k)}{\chi_s k^2}
 = 2 \frac{c_s(\omega_s) S_s(k)}{k}
 = c_s^2(\omega_s) \chi_s,
\end{equation}
where we are indicating by $c_s(\omega_s)$ the speed of sound in the spin
channel as extracted from the fit to Eq.~\eqref{eq:omega_decay}.
The data show fair agreement among the above expressions and between them and 
the direct measurement of $\rho - \rhodrag$. 
A notable exception are the data for $c_s^2[S_s(k)]/\chi_s$, where 
$c_s[S_s(k)]$ is the speed of sound in the spin channel, computed \emph{as if} 
Bijl-Feynman relation held.
We observe that this expression leads to the wrong behaviour in the region 
where $\rho-\rhodrag$ differs from one, i.e., where the drag effect is more 
prominent.

In both Figs~\ref{fig:dqmc:c} and~\ref{fig:dqmc}, the errobars in the
quantities extracted from the static structure factor and the winding number
come from statistical averaging. Spectral frequency, compressibility and
susceptibility have additional contributions to the error due to the use of
fitting procedures, Eqs.~\eqref{eq:susceptibility} and~\eqref{eq:omega_decay}.
The errors are effectively increased in some of the results, due to
cancellation of opposite trends as a function of $h$. The finite-size effects
are important for a quantitative agreement, as can be seen from
Fig.~\ref{fig:dqmc:c}a. It is not obvious that different quantities have
similar finite-size correction, which might eventually be responsible for some
remaining differences between various estimations in Fig.~\ref{fig:dqmc}.

\section{Dynamic stability}\label{sec:dynamic_stability}

Given the recent advances in measuring the spin superfluidity and its critical 
dynamics, we devote the final section to explore the consequences of the 
presence of a superfluid drag term on such phenomena. Both long living spin 
oscillations \cite{PhysRevA.94.063652} and critical spin superflow 
\cite{PhysRevLett.119.185302} for Bose-Bose mixtures as well as for Fermi-Bose 
superfluid mixtures \cite{PhysRevLett.115.265303} have been measured in the 
weakly interacting regime. The agreement with the available estimates 
\cite{ComptesRendusPhysique.16.241,EurPhysJD.69.126} is reasonable but rather 
far from being quantitative.

In the following, we determine the critical relative velocity required to
trigger the dynamical instability of a binary mixture of superfluids at zero
temperature. To this end, we generalise Eqs.~\eqref{eq:energy_Y}
and~\eqref{eq:Hhydro} and the results of Refs.~\cite{JLowTempPhys.150.612, 
JLowTempPhys.155.219, EurPhysJD.69.126} by considering the energy functional
\begin{align}
 E[&n_1, n_2, \phi_1, \phi_2] = \nonumber \\
 &= \int \left\lbrace \frac{\hbar^2}{2}
     \sum_\alpha  \left[ \frac{m_\alpha n_\alpha - \rhodrag}{m_\alpha^2}
     (\nabla \phi_\alpha)^2  \right]
     \right.
     \nonumber \\
     &\left. \quad
     +\frac{\rhodrag}{m_1 m_2} \nabla \phi_1 \cdot \nabla \phi_2 
  + e(n_1, n_2)\, \right\rbrace d^Dx
\end{align}
with $e(n_1, n_2)$ the internal energy density.
We subtracted $\rhodrag$ from the diagonal kinetic terms, so that the 
condition \eqref{eq:density_condition} on the total density is automatically 
satisfied.
Considering $n_\alpha$ and $\phi_\alpha$ as conjugate variables, the Hamilton
equations $m_\alpha \partial_t n_\alpha = \delta E / \delta \phi_\alpha$ and
$m_\alpha \partial_t \phi_\alpha = - \delta E / \delta n_\alpha$, yield
(implying $\alpha \neq \beta$)
\begin{align}
 & \partial_t n_\alpha =
       -\nabla(n_\alpha \vec{v}_\alpha)
       + m_\alpha^{-1}\nabla \left[\rhodrag (\vec{v}_\alpha - \vec{v}_\beta)
                             \right], \label{eq:ndot} \\
 & \partial_t \vec{v}_\alpha =
      - \nabla \frac{v_\alpha^2}{2}
      + \nabla \left[
            \frac{\eta_\alpha}{m_\alpha}
            \left( \frac{v_\alpha^2}{2} - \vec{v}_1 \cdot \vec{v}_2 \right)
          \right]
      - \frac{\nabla \mu_\alpha}{m_\alpha},
    \label{eq:vdot}
\end{align}
where $\vec{v}_\alpha = (\hbar / m) \nabla \phi_\alpha$ are the superfluid 
velocities, $\mu_\alpha = \partial \epsilon / \partial n_\alpha$ the chemical 
potentials and we defined
$\eta_\alpha \equiv \partial \rhodrag / \partial n_\alpha$, implicitly assuming
that $\rhodrag$ is a well-behaved function of the densities. The previous
system of hydrodynamic equations is satisfied by a steady state solution of
uniform velocity, matter and drag fields, such that all gradient terms vanish.
With a slight change of notation, we perturb about this solution by expanding
the density and velocity fields as
\begin{align}
 & n_\alpha \mapsto \bar{n}_\alpha + n_\alpha e^{i(\vec{qr} - \omega t)}, \\
 & \vec{v}_\alpha \mapsto \bar{\vec{v}}_\alpha + \vec{v}_\alpha
                                              e^{i(\vec{qr} - \omega t)},
\end{align}
and only keep the first order in the fluctuations, whereas, since it is
already small with respect to $n_\alpha m_\alpha$, we only keep the zero-th
order in $\rhodrag$.
Substituting these expansions into Eqs.~\eqref{eq:ndot}-\eqref{eq:vdot}, we
obtain
\begin{align}
\label{eq:n_expanded}
 & n_\alpha [\omega - \bar{\vec{v}}_\alpha \cdot \vec{q}] =
   \bar{n}_\alpha \vec{v}_\alpha \cdot \vec{q}
   - \frac{\rhodrag}{m_\alpha}
      (\vec{v}_\alpha - \vec{v}_\beta) \cdot \vec{q},\\
 & \vec{v}_\alpha [\omega - \bar{\vec{v}}_\alpha \cdot \vec{q}] =
   \frac{\vec{q}}{m_\alpha} \sum_{\gamma=1,2}
        \mu_{\alpha\gamma} n_\gamma,
\label{eq:v_expanded}
\end{align}
where we abbreviated $\mu_{\alpha \beta} \equiv \partial \mu_\alpha / \partial
\bar{n}_\beta$ (hence $\mu_{12} = \mu_{21}$, as expected by the symmetry of
interactions). In Eq.~\eqref{eq:v_expanded}, we neglected the term proportional
to $\eta_\alpha$, as the leading order is $O(\rhodrag n_\alpha, \rhodrag
v_\alpha)$, and the zero-th order, homogeneous by hypothesis, vanishes under
spatial differentiation.

Eliminating $n_\alpha$ and $v_\alpha$ in the system of equations
\eqref{eq:n_expanded}-\eqref{eq:v_expanded} leads to
\begin{widetext}
\begin{align}
& \left\lbrace \Omega_1^2 \Omega_2 -
 \left[ c_1^2 \Omega_2 -
 \rhodrag \left( \frac{\mu_{11}\Omega_2}{m_1^2} -
                   \frac{\mu_{21}\Omega_1}{m_1 m_2} \right) \right]
q^2 \right\rbrace
 \left\lbrace \Omega_2^2 \Omega_1 -
 \left[ c_2^2 \Omega_1 -
 \rhodrag \left( \frac{\mu_{22}\Omega_1}{m_2^2} -
                 \frac{\mu_{12}\Omega_2}{m_1 m_2} \right) \right]
q^2 \right\rbrace \nonumber
= \\
& \qquad \qquad \qquad \qquad \qquad \qquad = q^4
\left[ \bar{n}_1 \frac{\mu_{12} \Omega_2}{m_1}
- \rhodrag \left( \frac{\mu_{12} \Omega_2}{m_1^2} -
\frac{\mu_{22}\Omega_1}{m_1 m_2}
\right) \right]
\left[ \bar{n}_2 \frac{\mu_{21} \Omega_1}{m_2}
- \rhodrag \left( \frac{\mu_{21} \Omega_1}{m_2^2} -
\frac{\mu_{11}\Omega_2}{m_1 m_2}
\right) \right]
, \label{eq:stability}
\end{align}
\end{widetext}
where we set $\Omega_\alpha \equiv \omega - \bar{\vec{v}}_\alpha \cdot \vec{q}$
and $c_\alpha^2 \equiv \bar{n}_\alpha \mu_{\alpha \alpha} / m_\alpha$, the
latter being the speed of sound of a single superfluid. We also define
\begin{equation}
 c_{12}^4 \equiv \frac{\bar{n}_1 \bar{n}_2}{m_1 m_2} \mu_{12}^2,
\end{equation}
which has got the dimension of a speed and measures the strength of
inter-species contact interactions. By assuming a linear dispersion $\omega =
Cq$, the one above is a sixth order equation for the speed of sound $C$ of the
mixture, which determines its stability. In particular, the presence of complex
roots flags a dynamical instability. By setting $\rhodrag = 0$, 
Eq.~\eqref{eq:stability} reduces to the problem of the stability of a mixture
interacting only via contact interactions, which was studied in depth in
\cite{EurPhysJD.69.126}. 

This equation simplifies considerably when rewritten in a symmetric frame of
reference (SFR) in which $v/2 = \bar{\vec{v}}_1 \cdot \vec{q} = -
\bar{\vec{v}}_2 \cdot \vec{q}$, so that the projection of the relative velocity
of the two fluids along $\vec{q}$ is simply $v$. When $n_1 m_1 = n_2 m_2$, this
choice corresponds to the frame of reference of the centre of mass of the
system. A brief analysis of the case $v=0$ is reported in 
the Appendix~\ref{app:nonZ2}.

\subsection{Symmetric mixture}

\begin{figure}[tb!]
\begin{center}
 \includegraphics{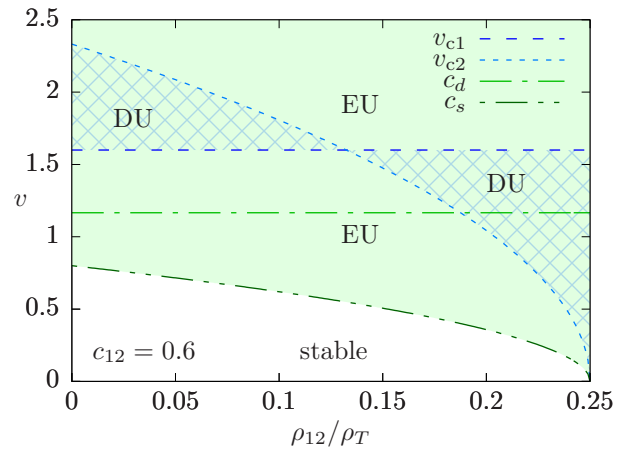}
\end{center}
 \caption{Stability diagram for a superfluid mixture with contact interactions
such that $c_{12} = 0.6$ and all speeds expressed in units of $c$. The spin
speed of sound $c_s$ is always below the density speed of sound $c_d$ (cf.\
Eq.~\eqref{eq:energetic_instability_thresholds}). The former delimits the
energetically unstable (EU) region (solid filling). The dynamically unstable
(DU) region (grid filling) lies between the critical velocities $v_{{\rm c}1}$
and $v_{{\rm c}2}$, cf.\ Eq.~\eqref{eq:vcrit_z2}. The threshold relative
velocities for both EU and DU vanish for $\rhodrag / \rho_T = 1/4$.}
\label{fig:stability}
\end{figure}

Some insight into the dynamic stability of the mixture can be gained by
considering a $\mathbb{Z}_2$ symmetric mixture, in which $\bar{n}_\alpha =
\bar{n}$, $m_\alpha = m$ and $\mu_{\alpha\alpha}$ are the same for both species
(hence also the speeds of sound coincide, $c_\alpha = c$). The stability
equation \eqref{eq:stability} then simplifies to a biquadratic equation whose
roots can be readily calculated. Restricting to positive relative velocities,
the condition that the two-fluid speed of sound be real then yields the
critical relative velocities for the stability of the mixture:
\begin{equation} \label{eq:vcrit_z2}
 \frac{v_{{\rm c}1}}{c} = 2 \sqrt{ 1 - \frac{c_{12}^2}{c^2}}, \quad
 \frac{v_{{\rm c}2}}{c} = 2 \sqrt{\left(1 +
\frac{c_{12}^2}{c^2}\right)\left(1-4\frac{\rhodrag}{\rho_T}\right)}.
\end{equation}
For $c_{12} = \rhodrag = 0$, the system is unstable for $v > 2 c$, confirming
the results of Refs.\ \cite{JLowTempPhys.150.612, JLowTempPhys.155.219,
ComptesRendusPhysique.16.241, EurPhysJD.69.126}.
When the relative velocity $v$ lies within $v_{{\rm c}1}$ and $v_{{\rm c}2}$,
the mixture becomes unstable, as schematically shown in Fig.\
\ref{fig:stability}.

It is worth reminding that the stability analysis presented here is valid 
within hydrodynamics, i.e., assuming a linear dispersion relation.
The inclusion of non-linear terms can give rise to finite momentum 
instabilities above the upper critical velocity, as it was pointed out in 
the case of a Bose-Bose mixture without superfluid drag 
\cite{PhysRevA.83.063602}.
Whether such finite momentum instability occurs in the same way also in 
presence of AB corrections is beyond the scope of the present paper and it 
will be discuss elsewhere. Nevertheless for large enough drag, 
\begin{equation}
 \frac{\rhodrag}{\rho_T} >
 \frac{1}{4} \left[ 1 - \frac{c^2 - c_{12}^2}{c^2 + c_{12}^2}\right].
\end{equation}
the lower critical velocity is $v_{{\rm c}2}$, which depends on the 
entrainment: in this regime, it should be possible observe a
shift in the onset of the dynamical instability. In particular for $\rhodrag =
\rho_T/4$, i.e., when the condition \eqref{eq:drag_bound} is saturated and the 
spin speed of sound vanishes, the mixture is unstable already at vanishing 
relative velocities.

For completeness, we compare the critical velocities for the dynamic stability 
\eqref{eq:vcrit_z2} with the speeds of sound of the density and spin modes,
Eqs.~\eqref{eq:cd}-\eqref{eq:cs}, respectively, which provide the critical 
velocity for the energetic (Landau) instability. With the notation of this 
section, they read
\begin{equation}\label{eq:energetic_instability_thresholds}
 c_d^2 = c^2 + c_{12}^2, \qquad
 c_s^2 = \left(1 - 4 \frac{\rhodrag}{\rho_T} \right) (c^2 - c_{12}^2),
\end{equation}
and they are also reported in Fig.~\ref{fig:stability}.
Although it occurs at a lower critical velocity, the energetic instability 
would not obfuscate the dynamical instability.
A counterflow experiment would only slightly trigger the Landau instability, 
which in any case would develop much more slowly than the dynamical one. 
Experimental methods based on the generation of soliton trains 
\cite{PhysRevA.84.041605, PhysRevLett.106.065302, PhysRevLett.119.185302} 
were shown to be able to cleanly identify dynamical instabilities of the kind 
discussed in this section.

\section{Conclusions} \label{sec:conclusions}

In the present work, we studied the physics of a superfluid mixture in the 
presence of current-current interactions, which lead to the 
so-called Andreev-Bashkin effect (AB).
Our minimal quantum hydrodynamic theory highlights the consequences of the 
presence of the superfluid drag $\rhodrag$ as an off-diagonal coefficient of 
the superfluid density matrix. In particular, for $\mathbb{Z}_2$ symmetric 
mixtures, in which the density and spin modes decouple, we predict the density 
channel to remain unaffected; the spin channel, on the other hand, exhibits 
corrections of linear order in $\rhodrag$ in both static quantities, 
such as the spin susceptibility $\chi_s$, and in dynamic quantities, such as 
the speed of sound of the spin mode, $c_s$. A prominent consequence of these 
corrections is that Bijl-Feynman theory, relating $c_s$ to $\chi_s$, is
no longer satisfied in the presence of AB corrections, as shown by 
Eq.~\eqref{eq:Ssk_vs_chi}.
In light of these results, we identify the speeds of sound and the 
susceptibilities as being the most promising observables in order to 
experimentally study the AB effect.
The quantum hydrodynamic theory further provides an upper bound to the value 
of $\rhodrag$, which must remain less than a fourth of the total superfluid 
density $\rho_T = \rho_1 + \rho_2 + 2\rhodrag$ (alternatively, in the language 
of the effective mass, $m^* \leq 2m$). The same bound appears also in the path 
integral formulation of superfluid densities, extended to a mixture of two 
superfluids, as well as a limiting case of the critical velocity 
\eqref{eq:vcrit_z2} for the dynamic stability of the mixture. 

The minimal toy model employed in Sec.~\ref{sec:quantum_hydro} also gives 
insight on low but finite temperature properties such as the specific heat, 
which is expected to depend on the drag through a factor $c_s^{-D}$ in $D$ 
dimensions. In general, finite temperature has a detrimental effect on the 
possibility of observing AB-related phenomena 
\cite{LowTempPhys.25.493,LowTempPhys.30.770}, due to the reduction of the 
superfluid density and to the emergence of dissipative drag between the two 
components. In this respect, the quantities we suggest to focus on to find 
experimental evidence of the presence of entrainment are much more suitable 
than trying to directly detect the current induced by one component on the 
other. It is worth mentioning that a recent experiment \cite{1708.03923} 
was able to record the dynamics of both superfluid and normal fractions of a 
weakly interacting Bose-Bose mixture, as well as the finite temperature 
polarisabilities. We believe that experiments such as this one pave the way 
towards the detection of subtle superfluid effects such as the AB drag.

As a check on the predictions of the quantum hydrodynamic model, we numerically 
investigated a system of $\mathbb{Z}_2$ symmetric bilayer dipolar bosons, with 
repulsive interactions within each layer and partially attractive interactions 
between the two layers.
The diffusion quantum Monte Carlo method allows the simultaneous measurement 
of several observables, separately on the density and spin channels. By 
inverting the theoretical relations between the static and dynamic observables, 
and the superfluid drag, we can compare indirect estimators of $\rhodrag$ with 
the direct estimator based on the winding numbers. Figure \ref{fig:dqmc} shows 
a good agreement between direct and indirect measurements of $(\rho - \rhodrag) 
/ nm$. A notable exception is when one extracts the speed of sound of the spin 
channel using the Bijl-Feynman relation, which, as aforementioned, breaks down 
if $\rhodrag \neq 0$.

We finally analysed the stability of the superfluid mixture. The condition for 
the onset of the (static) phase separation instability only involves the 
density channel, and hence is not modified by $\rhodrag$. On the other hand, 
when the two fluids are put in relative motion, for some values of the relative 
velocity the mixture becomes dynamically unstable. The critical velocities, 
reported in Eqs.~\eqref{eq:vcrit_z2}, carry a dependence on the superfluid 
drag. This result extends the previous ones \cite{JLowTempPhys.150.612, 
JLowTempPhys.155.219, EurPhysJD.69.126, ComptesRendusPhysique.16.241} which 
analysed a system with contact interactions only.

In the light of recent experimental works \cite{PhysRevLett.115.203201, 
PhysRevLett.115.265303, PhysRevLett.118.055301}, it is reasonable to claim 
that, albeit the AB is expected to be a subdominant effect, it may still be 
within reach of current experimental technology. It further opens up the game 
to interesting new phenomenological effects: for instance, if one is able to 
phase imprint a vortex on one species of the mixture, one may expect at least 
part of the vorticity to be transferred in a dissipationless fashion to the 
other species. This \emph{sympathetic stirring} of a superfluid mixture could 
become a useful tool in the study of vortex dynamics in superfluid mixtures. 
The stability conditions we put forward in Sec.~\ref{sec:dynamic_stability} 
and Appendix~\ref{app:nonZ2} may also have consequences in astrophysics, and 
in particular on the rotation profiles of neutron star cores 
\cite{AstrophysJ.836.203}.

\begin{acknowledgments}
We acknowledge useful discussions with Giovanni Barontini, Stefano Giorgini, 
Francesco Minardi, Lode Pollet and Sandro Stringari.
G.\ E.\ A.\ acknowledges partial financial support from the MICINN
(Spain) Grant No.~FIS2014-56257-C2-1-P.
The Barcelona Supercomputing Center (The Spanish National
Supercomputing Center -- Centro Nacional de Supercomputaci\'on) is
acknowledged for the computational facilities provided.
J.\ N.\ acknowledges support from the European Research Council through 
FP7/ERC Starting Grant No. 306897.
The authors gratefully acknowledge the Gauss Centre for Supercomputing
e.V.\ (\url{gauss-centre.eu}) for funding this project by providing
computing time on the GCS Supercomputer SuperMUC at Leibniz
Supercomputing Centre (LRZ, \url{lrz.de}).

\end{acknowledgments}

\bibliography{biblio}

\begin{thebibliography}{53}%
\makeatletter
\providecommand \@ifxundefined [1]{%
 \@ifx{#1\undefined}
}%
\providecommand \@ifnum [1]{%
 \ifnum #1\expandafter \@firstoftwo
 \else \expandafter \@secondoftwo
 \fi
}%
\providecommand \@ifx [1]{%
 \ifx #1\expandafter \@firstoftwo
 \else \expandafter \@secondoftwo
 \fi
}%
\providecommand \natexlab [1]{#1}%
\providecommand \enquote  [1]{``#1''}%
\providecommand \bibnamefont  [1]{#1}%
\providecommand \bibfnamefont [1]{#1}%
\providecommand \citenamefont [1]{#1}%
\providecommand \href@noop [0]{\@secondoftwo}%
\providecommand \href [0]{\begingroup \@sanitize@url \@href}%
\providecommand \@href[1]{\@@startlink{#1}\@@href}%
\providecommand \@@href[1]{\endgroup#1\@@endlink}%
\providecommand \@sanitize@url [0]{\catcode `\\12\catcode `\$12\catcode
  `\&12\catcode `\#12\catcode `\^12\catcode `\_12\catcode `\%12\relax}%
\providecommand \@@startlink[1]{}%
\providecommand \@@endlink[0]{}%
\providecommand \url  [0]{\begingroup\@sanitize@url \@url }%
\providecommand \@url [1]{\endgroup\@href {#1}{\urlprefix }}%
\providecommand \urlprefix  [0]{URL }%
\providecommand \Eprint [0]{\href }%
\providecommand \doibase [0]{http://dx.doi.org/}%
\providecommand \selectlanguage [0]{\@gobble}%
\providecommand \bibinfo  [0]{\@secondoftwo}%
\providecommand \bibfield  [0]{\@secondoftwo}%
\providecommand \translation [1]{[#1]}%
\providecommand \BibitemOpen [0]{}%
\providecommand \bibitemStop [0]{}%
\providecommand \bibitemNoStop [0]{.\EOS\space}%
\providecommand \EOS [0]{\spacefactor3000\relax}%
\providecommand \BibitemShut  [1]{\csname bibitem#1\endcsname}%
\let\auto@bib@innerbib\@empty
\bibitem [{\citenamefont {Edwards}\ \emph {et~al.}(1965)\citenamefont
  {Edwards}, \citenamefont {Brewer}, \citenamefont {Seligman}, \citenamefont
  {Skertic},\ and\ \citenamefont {Yaqub}}]{PhysRevLett.15.773}%
  \BibitemOpen
  \bibfield  {author} {\bibinfo {author} {\bibfnamefont {D.~O.}\ \bibnamefont
  {Edwards}}, \bibinfo {author} {\bibfnamefont {D.~F.}\ \bibnamefont {Brewer}},
  \bibinfo {author} {\bibfnamefont {P.}~\bibnamefont {Seligman}}, \bibinfo
  {author} {\bibfnamefont {M.}~\bibnamefont {Skertic}}, \ and\ \bibinfo
  {author} {\bibfnamefont {M.}~\bibnamefont {Yaqub}},\ }\bibfield  {title}
  {\enquote {\bibinfo {title} {Solubility of {${\mathrm{He}}^{3}$} in liquid
  {${\mathrm{He}}^{4}$} at {$0^{\circ}$K}},}\ }\href {\doibase
  10.1103/PhysRevLett.15.773} {\bibfield  {journal} {\bibinfo  {journal} {Phys.
  Rev. Lett.}\ }\textbf {\bibinfo {volume} {15}},\ \bibinfo {pages} {773--775}
  (\bibinfo {year} {1965})}\BibitemShut {NoStop}%
\bibitem [{\citenamefont {{Alpar}}\ \emph {et~al.}(1984)\citenamefont
  {{Alpar}}, \citenamefont {{Langer}},\ and\ \citenamefont
  {{Sauls}}}]{AstrophysJ.282.533}%
  \BibitemOpen
  \bibfield  {author} {\bibinfo {author} {\bibfnamefont {M.~A.}\ \bibnamefont
  {{Alpar}}}, \bibinfo {author} {\bibfnamefont {S.~A.}\ \bibnamefont
  {{Langer}}}, \ and\ \bibinfo {author} {\bibfnamefont {J.~A.}\ \bibnamefont
  {{Sauls}}},\ }\bibfield  {title} {\enquote {\bibinfo {title} {{Rapid
  postglitch spin-up of the superfluid core in pulsars}},}\ }\href {\doibase
  10.1086/162232} {\bibfield  {journal} {\bibinfo  {journal} {\apj}\ }\textbf
  {\bibinfo {volume} {282}},\ \bibinfo {pages} {533--541} (\bibinfo {year}
  {1984})}\BibitemShut {NoStop}%
\bibitem [{\citenamefont {Kobyakov}\ and\ \citenamefont
  {Pethick}(2017)}]{AstrophysJ.836.203}%
  \BibitemOpen
  \bibfield  {author} {\bibinfo {author} {\bibfnamefont {D.~N.}\ \bibnamefont
  {Kobyakov}}\ and\ \bibinfo {author} {\bibfnamefont {C.~J.}\ \bibnamefont
  {Pethick}},\ }\bibfield  {title} {\enquote {\bibinfo {title} {Two-component
  superfluid hydrodynamics of neutron star cores},}\ }\href
  {http://stacks.iop.org/0004-637X/836/i=2/a=203} {\bibfield  {journal}
  {\bibinfo  {journal} {Astrophys. J.}\ }\textbf {\bibinfo {volume} {836}},\
  \bibinfo {pages} {203} (\bibinfo {year} {2017})}\BibitemShut {NoStop}%
\bibitem [{\citenamefont {Garaud}\ \emph {et~al.}(2014)\citenamefont {Garaud},
  \citenamefont {Sellin}, \citenamefont {J\"aykk\"a},\ and\ \citenamefont
  {Babaev}}]{PhysRevB.89.104508}%
  \BibitemOpen
  \bibfield  {author} {\bibinfo {author} {\bibfnamefont {Julien}\ \bibnamefont
  {Garaud}}, \bibinfo {author} {\bibfnamefont {Karl A.~H.}\ \bibnamefont
  {Sellin}}, \bibinfo {author} {\bibfnamefont {Juha}\ \bibnamefont
  {J\"aykk\"a}}, \ and\ \bibinfo {author} {\bibfnamefont {Egor}\ \bibnamefont
  {Babaev}},\ }\bibfield  {title} {\enquote {\bibinfo {title} {Skyrmions
  induced by dissipationless drag in
  {U(1)$\ifmmode\times\else\texttimes\fi{}$U(1)} superconductors},}\ }\href
  {\doibase 10.1103/PhysRevB.89.104508} {\bibfield  {journal} {\bibinfo
  {journal} {Phys. Rev. B}\ }\textbf {\bibinfo {volume} {89}},\ \bibinfo
  {pages} {104508} (\bibinfo {year} {2014})}\BibitemShut {NoStop}%
\bibitem [{\citenamefont {Modugno}\ \emph {et~al.}(2002)\citenamefont
  {Modugno}, \citenamefont {Modugno}, \citenamefont {Riboli}, \citenamefont
  {Roati},\ and\ \citenamefont {Inguscio}}]{PhysRevLett.89.190404}%
  \BibitemOpen
  \bibfield  {author} {\bibinfo {author} {\bibfnamefont {G.}~\bibnamefont
  {Modugno}}, \bibinfo {author} {\bibfnamefont {M.}~\bibnamefont {Modugno}},
  \bibinfo {author} {\bibfnamefont {F.}~\bibnamefont {Riboli}}, \bibinfo
  {author} {\bibfnamefont {G.}~\bibnamefont {Roati}}, \ and\ \bibinfo {author}
  {\bibfnamefont {M.}~\bibnamefont {Inguscio}},\ }\bibfield  {title} {\enquote
  {\bibinfo {title} {Two atomic species superfluid},}\ }\href {\doibase
  10.1103/PhysRevLett.89.190404} {\bibfield  {journal} {\bibinfo  {journal}
  {Phys. Rev. Lett.}\ }\textbf {\bibinfo {volume} {89}},\ \bibinfo {pages}
  {190404} (\bibinfo {year} {2002})}\BibitemShut {NoStop}%
\bibitem [{\citenamefont {Catani}\ \emph {et~al.}(2008)\citenamefont {Catani},
  \citenamefont {De~Sarlo}, \citenamefont {Barontini}, \citenamefont
  {Minardi},\ and\ \citenamefont {Inguscio}}]{PhysRevA.77.011603}%
  \BibitemOpen
  \bibfield  {author} {\bibinfo {author} {\bibfnamefont {J.}~\bibnamefont
  {Catani}}, \bibinfo {author} {\bibfnamefont {L.}~\bibnamefont {De~Sarlo}},
  \bibinfo {author} {\bibfnamefont {G.}~\bibnamefont {Barontini}}, \bibinfo
  {author} {\bibfnamefont {F.}~\bibnamefont {Minardi}}, \ and\ \bibinfo
  {author} {\bibfnamefont {M.}~\bibnamefont {Inguscio}},\ }\bibfield  {title}
  {\enquote {\bibinfo {title} {Degenerate {Bose-Bose} mixture in a
  three-dimensional optical lattice},}\ }\href {\doibase
  10.1103/PhysRevA.77.011603} {\bibfield  {journal} {\bibinfo  {journal} {Phys.
  Rev. A}\ }\textbf {\bibinfo {volume} {77}},\ \bibinfo {pages} {011603}
  (\bibinfo {year} {2008})}\BibitemShut {NoStop}%
\bibitem [{\citenamefont {Roy}\ \emph {et~al.}(2017)\citenamefont {Roy},
  \citenamefont {Green}, \citenamefont {Bowler},\ and\ \citenamefont
  {Gupta}}]{PhysRevLett.118.055301}%
  \BibitemOpen
  \bibfield  {author} {\bibinfo {author} {\bibfnamefont {Richard}\ \bibnamefont
  {Roy}}, \bibinfo {author} {\bibfnamefont {Alaina}\ \bibnamefont {Green}},
  \bibinfo {author} {\bibfnamefont {Ryan}\ \bibnamefont {Bowler}}, \ and\
  \bibinfo {author} {\bibfnamefont {Subhadeep}\ \bibnamefont {Gupta}},\
  }\bibfield  {title} {\enquote {\bibinfo {title} {Two-element mixture of
  {Bose} and {Fermi} superfluids},}\ }\href {\doibase
  10.1103/PhysRevLett.118.055301} {\bibfield  {journal} {\bibinfo  {journal}
  {Phys. Rev. Lett.}\ }\textbf {\bibinfo {volume} {118}},\ \bibinfo {pages}
  {055301} (\bibinfo {year} {2017})}\BibitemShut {NoStop}%
\bibitem [{\citenamefont {Lagoudakis}\ \emph {et~al.}(2009)\citenamefont
  {Lagoudakis}, \citenamefont {Ostatnick{\'y}}, \citenamefont {Kavokin},
  \citenamefont {Rubo}, \citenamefont {Andr{\'e}},\ and\ \citenamefont
  {Deveaud-Pl{\'e}dran}}]{Science.326.974}%
  \BibitemOpen
  \bibfield  {author} {\bibinfo {author} {\bibfnamefont {K.~G.}\ \bibnamefont
  {Lagoudakis}}, \bibinfo {author} {\bibfnamefont {T.}~\bibnamefont
  {Ostatnick{\'y}}}, \bibinfo {author} {\bibfnamefont {A.~V.}\ \bibnamefont
  {Kavokin}}, \bibinfo {author} {\bibfnamefont {Y.~G.}\ \bibnamefont {Rubo}},
  \bibinfo {author} {\bibfnamefont {R.}~\bibnamefont {Andr{\'e}}}, \ and\
  \bibinfo {author} {\bibfnamefont {B.}~\bibnamefont {Deveaud-Pl{\'e}dran}},\
  }\bibfield  {title} {\enquote {\bibinfo {title} {Observation of half-quantum
  vortices in an exciton-polariton condensate},}\ }\href {\doibase
  10.1126/science.1177980} {\bibfield  {journal} {\bibinfo  {journal}
  {Science}\ }\textbf {\bibinfo {volume} {326}},\ \bibinfo {pages} {974--976}
  (\bibinfo {year} {2009})}\BibitemShut {NoStop}%
\bibitem [{\citenamefont {{Andreev}}\ and\ \citenamefont
  {{Bashkin}}(1976)}]{JETP.42.164}%
  \BibitemOpen
  \bibfield  {author} {\bibinfo {author} {\bibfnamefont {A.~F.}\ \bibnamefont
  {{Andreev}}}\ and\ \bibinfo {author} {\bibfnamefont {E.~P.}\ \bibnamefont
  {{Bashkin}}},\ }\bibfield  {title} {\enquote {\bibinfo {title}
  {{Three-velocity hydrodynamics of superfluid solutions}},}\ }\href
  {http://jetp.ac.ru/cgi-bin/e/index/e/42/1/p164?a=list} {\bibfield  {journal}
  {\bibinfo  {journal} {Sov. Phys.-JETP}\ }\textbf {\bibinfo {volume} {42}},\
  \bibinfo {pages} {164--167} (\bibinfo {year} {1976})}\BibitemShut {NoStop}%
\bibitem [{\citenamefont {Khalatnikov}(1957)}]{JETP.5.542}%
  \BibitemOpen
  \bibfield  {author} {\bibinfo {author} {\bibfnamefont {I.~M.}\ \bibnamefont
  {Khalatnikov}},\ }\bibfield  {title} {\enquote {\bibinfo {title}
  {Hydrodynamics of solutions of 2 superfluid liquids},}\ }\href
  {http://jetp.ac.ru/cgi-bin/e/index/e/5/4/p542?a=list} {\bibfield  {journal}
  {\bibinfo  {journal} {Sov. Phys.-JETP}\ }\textbf {\bibinfo {volume} {5}},\
  \bibinfo {pages} {542--545} (\bibinfo {year} {1957})}\BibitemShut {NoStop}%
\bibitem [{\citenamefont {Mineev}(1975)}]{JETP.40.338}%
  \BibitemOpen
  \bibfield  {author} {\bibinfo {author} {\bibfnamefont {V.~P.}\ \bibnamefont
  {Mineev}},\ }\bibfield  {title} {\enquote {\bibinfo {title} {Some problems in
  the hydrodynamics of solutions of two superfluid liquids},}\ }\href
  {http://jetp.ac.ru/cgi-bin/e/index/e/40/2/p338?a=list} {\bibfield  {journal}
  {\bibinfo  {journal} {Sov. Phys.-JETP}\ }\textbf {\bibinfo {volume} {40}},\
  \bibinfo {pages} {338--341} (\bibinfo {year} {1975})}\BibitemShut {NoStop}%
\bibitem [{\citenamefont {Meyerovich}(1984)}]{JETP.60.741}%
  \BibitemOpen
  \bibfield  {author} {\bibinfo {author} {\bibfnamefont {A.~E.}\ \bibnamefont
  {Meyerovich}},\ }\bibfield  {title} {\enquote {\bibinfo {title} {Dynamics of
  superfluid {$^3$He} in {$^3$He-$^4$He} solutions},}\ }\href
  {http://jetp.ac.ru/cgi-bin/e/index/e/60/4/p741?a=list} {\bibfield  {journal}
  {\bibinfo  {journal} {Sov. Phys.-JETP}\ }\textbf {\bibinfo {volume} {60}},\
  \bibinfo {pages} {741--747} (\bibinfo {year} {1984})}\BibitemShut {NoStop}%
\bibitem [{\citenamefont {Lattimer}\ and\ \citenamefont
  {Prakash}(2004)}]{Lattimer536}%
  \BibitemOpen
  \bibfield  {author} {\bibinfo {author} {\bibfnamefont {J.~M.}\ \bibnamefont
  {Lattimer}}\ and\ \bibinfo {author} {\bibfnamefont {M.}~\bibnamefont
  {Prakash}},\ }\bibfield  {title} {\enquote {\bibinfo {title} {The physics of
  neutron stars},}\ }\href {\doibase 10.1126/science.1090720} {\bibfield
  {journal} {\bibinfo  {journal} {Science}\ }\textbf {\bibinfo {volume}
  {304}},\ \bibinfo {pages} {536--542} (\bibinfo {year} {2004})}\BibitemShut
  {NoStop}%
\bibitem [{\citenamefont {Babaev}(2004)}]{PhysRevD.70.043001}%
  \BibitemOpen
  \bibfield  {author} {\bibinfo {author} {\bibfnamefont {Egor}\ \bibnamefont
  {Babaev}},\ }\bibfield  {title} {\enquote {\bibinfo {title}
  {{Andreev-Bashkin} effect and knot solitons in an interacting mixture of a
  charged and a neutral superfluid with possible relevance for neutron
  stars},}\ }\href {\doibase 10.1103/PhysRevD.70.043001} {\bibfield  {journal}
  {\bibinfo  {journal} {Phys. Rev. D}\ }\textbf {\bibinfo {volume} {70}},\
  \bibinfo {pages} {043001} (\bibinfo {year} {2004})}\BibitemShut {NoStop}%
\bibitem [{\citenamefont {Fil}\ and\ \citenamefont
  {Shevchenko}(2004)}]{LowTempPhys.30.770}%
  \BibitemOpen
  \bibfield  {author} {\bibinfo {author} {\bibfnamefont {D.~V.}\ \bibnamefont
  {Fil}}\ and\ \bibinfo {author} {\bibfnamefont {S.~I.}\ \bibnamefont
  {Shevchenko}},\ }\bibfield  {title} {\enquote {\bibinfo {title} {Drag of
  superfluid current in bilayer {Bose} systems},}\ }\href {\doibase
  10.1063/1.1808194} {\bibfield  {journal} {\bibinfo  {journal} {Low Temp.
  Phys.}\ }\textbf {\bibinfo {volume} {30}},\ \bibinfo {pages} {770--777}
  (\bibinfo {year} {2004})}\BibitemShut {NoStop}%
\bibitem [{\citenamefont {Fil}\ and\ \citenamefont
  {Shevchenko}(2005)}]{PhysRevA.72.013616}%
  \BibitemOpen
  \bibfield  {author} {\bibinfo {author} {\bibfnamefont {D.~V.}\ \bibnamefont
  {Fil}}\ and\ \bibinfo {author} {\bibfnamefont {S.~I.}\ \bibnamefont
  {Shevchenko}},\ }\bibfield  {title} {\enquote {\bibinfo {title}
  {Nondissipative drag of superflow in a two-component {Bose} gas},}\ }\href
  {\doibase 10.1103/PhysRevA.72.013616} {\bibfield  {journal} {\bibinfo
  {journal} {Phys. Rev. A}\ }\textbf {\bibinfo {volume} {72}},\ \bibinfo
  {pages} {013616} (\bibinfo {year} {2005})}\BibitemShut {NoStop}%
\bibitem [{\citenamefont {Linder}\ and\ \citenamefont
  {Sudb\o{}}(2009)}]{PhysRevA.79.063610}%
  \BibitemOpen
  \bibfield  {author} {\bibinfo {author} {\bibfnamefont {Jacob}\ \bibnamefont
  {Linder}}\ and\ \bibinfo {author} {\bibfnamefont {Asle}\ \bibnamefont
  {Sudb\o{}}},\ }\bibfield  {title} {\enquote {\bibinfo {title} {Calculation of
  drag and superfluid velocity from the microscopic parameters and excitation
  energies of a two-component {Bose-Einstein} condensate in an optical
  lattice},}\ }\href {\doibase 10.1103/PhysRevA.79.063610} {\bibfield
  {journal} {\bibinfo  {journal} {Phys. Rev. A}\ }\textbf {\bibinfo {volume}
  {79}},\ \bibinfo {pages} {063610} (\bibinfo {year} {2009})}\BibitemShut
  {NoStop}%
\bibitem [{\citenamefont {Hofer}\ \emph {et~al.}(2012)\citenamefont {Hofer},
  \citenamefont {Bruder},\ and\ \citenamefont {Stojanovi\ifmmode~\acute{c}\else
  \'{c}\fi{}}}]{PhysRevA.86.033627}%
  \BibitemOpen
  \bibfield  {author} {\bibinfo {author} {\bibfnamefont {Patrick~P.}\
  \bibnamefont {Hofer}}, \bibinfo {author} {\bibfnamefont {C.}~\bibnamefont
  {Bruder}}, \ and\ \bibinfo {author} {\bibfnamefont {Vladimir~M.}\
  \bibnamefont {Stojanovi\ifmmode~\acute{c}\else \'{c}\fi{}}},\ }\bibfield
  {title} {\enquote {\bibinfo {title} {Superfluid drag of two-species
  {Bose-Einstein} condensates in optical lattices},}\ }\href {\doibase
  10.1103/PhysRevA.86.033627} {\bibfield  {journal} {\bibinfo  {journal} {Phys.
  Rev. A}\ }\textbf {\bibinfo {volume} {86}},\ \bibinfo {pages} {033627}
  (\bibinfo {year} {2012})}\BibitemShut {NoStop}%
\bibitem [{\citenamefont {Leggett}(1998)}]{Leggett1998}%
  \BibitemOpen
  \bibfield  {author} {\bibinfo {author} {\bibfnamefont {A.~J.}\ \bibnamefont
  {Leggett}},\ }\bibfield  {title} {\enquote {\bibinfo {title} {On the
  superfluid fraction of an arbitrary many-body system at {$T=0$}},}\ }\href
  {\doibase 10.1023/B:JOSS.0000033170.38619.6c} {\bibfield  {journal} {\bibinfo
   {journal} {J. Stat. Phys.}\ }\textbf {\bibinfo {volume} {93}},\ \bibinfo
  {pages} {927--941} (\bibinfo {year} {1998})}\BibitemShut {NoStop}%
\bibitem [{\citenamefont {Beattie}\ \emph {et~al.}(2013)\citenamefont
  {Beattie}, \citenamefont {Moulder}, \citenamefont {Fletcher},\ and\
  \citenamefont {Hadzibabic}}]{ZoranCurrent}%
  \BibitemOpen
  \bibfield  {author} {\bibinfo {author} {\bibfnamefont {Scott}\ \bibnamefont
  {Beattie}}, \bibinfo {author} {\bibfnamefont {Stuart}\ \bibnamefont
  {Moulder}}, \bibinfo {author} {\bibfnamefont {Richard~J.}\ \bibnamefont
  {Fletcher}}, \ and\ \bibinfo {author} {\bibfnamefont {Zoran}\ \bibnamefont
  {Hadzibabic}},\ }\bibfield  {title} {\enquote {\bibinfo {title} {Persistent
  currents in spinor condensates},}\ }\href {\doibase
  10.1103/PhysRevLett.110.025301} {\bibfield  {journal} {\bibinfo  {journal}
  {Phys. Rev. Lett.}\ }\textbf {\bibinfo {volume} {110}},\ \bibinfo {pages}
  {025301} (\bibinfo {year} {2013})}\BibitemShut {NoStop}%
\bibitem [{\citenamefont {Abad}\ \emph {et~al.}(2014)\citenamefont {Abad},
  \citenamefont {Sartori}, \citenamefont {Finazzi},\ and\ \citenamefont
  {Recati}}]{AlessioCurrent}%
  \BibitemOpen
  \bibfield  {author} {\bibinfo {author} {\bibfnamefont {M.}~\bibnamefont
  {Abad}}, \bibinfo {author} {\bibfnamefont {A.}~\bibnamefont {Sartori}},
  \bibinfo {author} {\bibfnamefont {S.}~\bibnamefont {Finazzi}}, \ and\
  \bibinfo {author} {\bibfnamefont {A.}~\bibnamefont {Recati}},\ }\bibfield
  {title} {\enquote {\bibinfo {title} {Persistent currents in two-component
  condensates in a toroidal trap},}\ }\href {\doibase
  10.1103/PhysRevA.89.053602} {\bibfield  {journal} {\bibinfo  {journal} {Phys.
  Rev. A}\ }\textbf {\bibinfo {volume} {89}},\ \bibinfo {pages} {053602}
  (\bibinfo {year} {2014})}\BibitemShut {NoStop}%
\bibitem [{\citenamefont {Kaurov}\ \emph {et~al.}(2005)\citenamefont {Kaurov},
  \citenamefont {Kuklov},\ and\ \citenamefont
  {Meyerovich}}]{PhysRevLett.95.090403}%
  \BibitemOpen
  \bibfield  {author} {\bibinfo {author} {\bibfnamefont {V.~M.}\ \bibnamefont
  {Kaurov}}, \bibinfo {author} {\bibfnamefont {A.~B.}\ \bibnamefont {Kuklov}},
  \ and\ \bibinfo {author} {\bibfnamefont {A.~E.}\ \bibnamefont {Meyerovich}},\
  }\bibfield  {title} {\enquote {\bibinfo {title} {Drag effect and topological
  complexes in strongly interacting two-component lattice superfluids},}\
  }\href {\doibase 10.1103/PhysRevLett.95.090403} {\bibfield  {journal}
  {\bibinfo  {journal} {Phys. Rev. Lett.}\ }\textbf {\bibinfo {volume} {95}},\
  \bibinfo {pages} {090403} (\bibinfo {year} {2005})}\BibitemShut {NoStop}%
\bibitem [{\citenamefont {Giuliani}\ and\ \citenamefont
  {Vignale}(2005)}]{GiulianiVignale}%
  \BibitemOpen
  \bibfield  {author} {\bibinfo {author} {\bibfnamefont {Gabriele}\
  \bibnamefont {Giuliani}}\ and\ \bibinfo {author} {\bibfnamefont {Giovanni}\
  \bibnamefont {Vignale}},\ }\href@noop {} {\emph {\bibinfo {title} {Quantum
  theory of the electron liquid}}}\ (\bibinfo  {publisher} {Cambridge
  University Press},\ \bibinfo {year} {2005})\BibitemShut {NoStop}%
\bibitem [{\citenamefont {Giamarchi}(2004)}]{Giamarchi}%
  \BibitemOpen
  \bibfield  {author} {\bibinfo {author} {\bibfnamefont {Thierry}\ \bibnamefont
  {Giamarchi}},\ }\href@noop {} {\emph {\bibinfo {title} {Quantum Physics in
  One Dimension}}}\ (\bibinfo  {publisher} {Oxford University Press},\ \bibinfo
  {year} {2004})\BibitemShut {NoStop}%
\bibitem [{\citenamefont {Schulz}\ \emph {et~al.}(1998)\citenamefont {Schulz},
  \citenamefont {Cuniberti},\ and\ \citenamefont {Pieri}}]{condmat.9807366}%
  \BibitemOpen
  \bibfield  {author} {\bibinfo {author} {\bibfnamefont {H.~J.}\ \bibnamefont
  {Schulz}}, \bibinfo {author} {\bibfnamefont {G.}~\bibnamefont {Cuniberti}}, \
  and\ \bibinfo {author} {\bibfnamefont {P.}~\bibnamefont {Pieri}},\ }\bibfield
   {title} {\enquote {\bibinfo {title} {{Fermi} liquids and {Luttinger}
  liquids},}\ }\href@noop {} {\  (\bibinfo {year} {1998})},\ \Eprint
  {http://arxiv.org/abs/cond-mat/9807366} {arXiv:cond-mat/9807366} \BibitemShut
  {NoStop}%
\bibitem [{\citenamefont {Kleine}\ \emph
  {et~al.}(2008{\natexlab{a}})\citenamefont {Kleine}, \citenamefont {Kollath},
  \citenamefont {McCulloch}, \citenamefont {Giamarchi},\ and\ \citenamefont
  {Schollw\"ock}}]{NewJPhys.10.045025}%
  \BibitemOpen
  \bibfield  {author} {\bibinfo {author} {\bibfnamefont {A}~\bibnamefont
  {Kleine}}, \bibinfo {author} {\bibfnamefont {C}~\bibnamefont {Kollath}},
  \bibinfo {author} {\bibfnamefont {I~P}\ \bibnamefont {McCulloch}}, \bibinfo
  {author} {\bibfnamefont {T}~\bibnamefont {Giamarchi}}, \ and\ \bibinfo
  {author} {\bibfnamefont {U}~\bibnamefont {Schollw\"ock}},\ }\bibfield
  {title} {\enquote {\bibinfo {title} {Excitations in two-component {Bose}
  gases},}\ }\href {http://stacks.iop.org/1367-2630/10/i=4/a=045025} {\bibfield
   {journal} {\bibinfo  {journal} {New J. Phys.}\ }\textbf {\bibinfo {volume}
  {10}},\ \bibinfo {pages} {045025} (\bibinfo {year}
  {2008}{\natexlab{a}})}\BibitemShut {NoStop}%
\bibitem [{\citenamefont {Kleine}\ \emph
  {et~al.}(2008{\natexlab{b}})\citenamefont {Kleine}, \citenamefont {Kollath},
  \citenamefont {McCulloch}, \citenamefont {Giamarchi},\ and\ \citenamefont
  {Schollw\"ock}}]{PhysRevA.77.013607}%
  \BibitemOpen
  \bibfield  {author} {\bibinfo {author} {\bibfnamefont {A.}~\bibnamefont
  {Kleine}}, \bibinfo {author} {\bibfnamefont {C.}~\bibnamefont {Kollath}},
  \bibinfo {author} {\bibfnamefont {I.~P.}\ \bibnamefont {McCulloch}}, \bibinfo
  {author} {\bibfnamefont {T.}~\bibnamefont {Giamarchi}}, \ and\ \bibinfo
  {author} {\bibfnamefont {U.}~\bibnamefont {Schollw\"ock}},\ }\bibfield
  {title} {\enquote {\bibinfo {title} {Spin-charge separation in two-component
  {Bose} gases},}\ }\href {\doibase 10.1103/PhysRevA.77.013607} {\bibfield
  {journal} {\bibinfo  {journal} {Phys. Rev. A}\ }\textbf {\bibinfo {volume}
  {77}},\ \bibinfo {pages} {013607} (\bibinfo {year}
  {2008}{\natexlab{b}})}\BibitemShut {NoStop}%
\bibitem [{\citenamefont {Parisi}\ and\ \citenamefont
  {Giorgini}()}]{parisi_giorgini_private}%
  \BibitemOpen
  \bibfield  {author} {\bibinfo {author} {\bibfnamefont {L.}~\bibnamefont
  {Parisi}}\ and\ \bibinfo {author} {\bibfnamefont {S.}~\bibnamefont
  {Giorgini}},\ }\href@noop {} {\bibinfo  {journal} {private communication}\
  }\BibitemShut {NoStop}%
\bibitem [{\citenamefont {Pollock}\ and\ \citenamefont
  {Ceperley}(1987)}]{PhysRevB.36.8343}%
  \BibitemOpen
\bibfield  {journal} {  }\bibfield  {author} {\bibinfo {author} {\bibfnamefont
  {E.~L.}\ \bibnamefont {Pollock}}\ and\ \bibinfo {author} {\bibfnamefont
  {D.~M.}\ \bibnamefont {Ceperley}},\ }\bibfield  {title} {\enquote {\bibinfo
  {title} {Path-integral computation of superfluid densities},}\ }\href
  {\doibase 10.1103/PhysRevB.36.8343} {\bibfield  {journal} {\bibinfo
  {journal} {Phys. Rev. B}\ }\textbf {\bibinfo {volume} {36}},\ \bibinfo
  {pages} {8343--8352} (\bibinfo {year} {1987})}\BibitemShut {NoStop}%
\bibitem [{\citenamefont {Terentjev}\ and\ \citenamefont
  {Shevchenko}(1999)}]{LowTempPhys.25.493}%
  \BibitemOpen
  \bibfield  {author} {\bibinfo {author} {\bibfnamefont {S.~V.}\ \bibnamefont
  {Terentjev}}\ and\ \bibinfo {author} {\bibfnamefont {S.~I.}\ \bibnamefont
  {Shevchenko}},\ }\bibfield  {title} {\enquote {\bibinfo {title} {On transfer
  of motion in a system of two-dimensional superfluid {Bose}-gases separated by
  a thin layer},}\ }\href {\doibase 10.1063/1.593773} {\bibfield  {journal}
  {\bibinfo  {journal} {Low Temp. Phys.}\ }\textbf {\bibinfo {volume} {25}},\
  \bibinfo {pages} {493--502} (\bibinfo {year} {1999})}\BibitemShut {NoStop}%
\bibitem [{\citenamefont {Ni}\ \emph {et~al.}(2010)\citenamefont {Ni},
  \citenamefont {Ospelkaus}, \citenamefont {Wang}, \citenamefont {Quemener},
  \citenamefont {Neyenhuis}, \citenamefont {de~Miranda}, \citenamefont {Bohn},
  \citenamefont {Ye},\ and\ \citenamefont {Jin}}]{Ni2010}%
  \BibitemOpen
  \bibfield  {author} {\bibinfo {author} {\bibfnamefont {K.-K.}\ \bibnamefont
  {Ni}}, \bibinfo {author} {\bibfnamefont {S.}~\bibnamefont {Ospelkaus}},
  \bibinfo {author} {\bibfnamefont {D.}~\bibnamefont {Wang}}, \bibinfo {author}
  {\bibfnamefont {G.}~\bibnamefont {Quemener}}, \bibinfo {author}
  {\bibfnamefont {B.}~\bibnamefont {Neyenhuis}}, \bibinfo {author}
  {\bibfnamefont {M.~H.~G.}\ \bibnamefont {de~Miranda}}, \bibinfo {author}
  {\bibfnamefont {J.~L.}\ \bibnamefont {Bohn}}, \bibinfo {author}
  {\bibfnamefont {J.}~\bibnamefont {Ye}}, \ and\ \bibinfo {author}
  {\bibfnamefont {D.}~\bibnamefont {Jin}},\ }\bibfield  {title} {\enquote
  {\bibinfo {title} {Dipolar collisions of polar molecules in the quantum
  regime},}\ }\href {\doibase doi:10.1038/nature08953} {\bibfield  {journal}
  {\bibinfo  {journal} {Nature}\ }\textbf {\bibinfo {volume} {464}},\ \bibinfo
  {pages} {1324--1328} (\bibinfo {year} {2010})}\BibitemShut {NoStop}%
\bibitem [{\citenamefont {Macia}\ \emph {et~al.}(2014)\citenamefont {Macia},
  \citenamefont {Astrakharchik}, \citenamefont {Mazzanti}, \citenamefont
  {Giorgini},\ and\ \citenamefont {Boronat}}]{PhysRevA.90.043623}%
  \BibitemOpen
  \bibfield  {author} {\bibinfo {author} {\bibfnamefont {A.}~\bibnamefont
  {Macia}}, \bibinfo {author} {\bibfnamefont {G.~E.}\ \bibnamefont
  {Astrakharchik}}, \bibinfo {author} {\bibfnamefont {F.}~\bibnamefont
  {Mazzanti}}, \bibinfo {author} {\bibfnamefont {S.}~\bibnamefont {Giorgini}},
  \ and\ \bibinfo {author} {\bibfnamefont {J.}~\bibnamefont {Boronat}},\
  }\bibfield  {title} {\enquote {\bibinfo {title} {Single-particle versus pair
  superfluidity in a bilayer system of dipolar bosons},}\ }\href {\doibase
  10.1103/PhysRevA.90.043623} {\bibfield  {journal} {\bibinfo  {journal} {Phys.
  Rev. A}\ }\textbf {\bibinfo {volume} {90}},\ \bibinfo {pages} {043623}
  (\bibinfo {year} {2014})}\BibitemShut {NoStop}%
\bibitem [{\citenamefont {Astrakharchik}\ \emph {et~al.}(2016)\citenamefont
  {Astrakharchik}, \citenamefont {Zillich}, \citenamefont {Mazzanti},\ and\
  \citenamefont {Boronat}}]{PhysRevA.94.063630}%
  \BibitemOpen
  \bibfield  {author} {\bibinfo {author} {\bibfnamefont {G.~E.}\ \bibnamefont
  {Astrakharchik}}, \bibinfo {author} {\bibfnamefont {R.~E.}\ \bibnamefont
  {Zillich}}, \bibinfo {author} {\bibfnamefont {F.}~\bibnamefont {Mazzanti}}, \
  and\ \bibinfo {author} {\bibfnamefont {J.}~\bibnamefont {Boronat}},\
  }\bibfield  {title} {\enquote {\bibinfo {title} {Gapped spectrum in
  pair-superfluid bosons},}\ }\href {\doibase 10.1103/PhysRevA.94.063630}
  {\bibfield  {journal} {\bibinfo  {journal} {Phys. Rev. A}\ }\textbf {\bibinfo
  {volume} {94}},\ \bibinfo {pages} {063630} (\bibinfo {year}
  {2016})}\BibitemShut {NoStop}%
\bibitem [{\citenamefont {Klawunn}\ \emph {et~al.}(2010)\citenamefont
  {Klawunn}, \citenamefont {Pikovski},\ and\ \citenamefont
  {Santos}}]{PhysRevA.82.044701}%
  \BibitemOpen
  \bibfield  {author} {\bibinfo {author} {\bibfnamefont {Michael}\ \bibnamefont
  {Klawunn}}, \bibinfo {author} {\bibfnamefont {Alexander}\ \bibnamefont
  {Pikovski}}, \ and\ \bibinfo {author} {\bibfnamefont {Luis}\ \bibnamefont
  {Santos}},\ }\bibfield  {title} {\enquote {\bibinfo {title} {Two-dimensional
  scattering and bound states of polar molecules in bilayers},}\ }\href
  {\doibase 10.1103/PhysRevA.82.044701} {\bibfield  {journal} {\bibinfo
  {journal} {Phys. Rev. A}\ }\textbf {\bibinfo {volume} {82}},\ \bibinfo
  {pages} {044701} (\bibinfo {year} {2010})}\BibitemShut {NoStop}%
\bibitem [{\citenamefont {Frisch}\ \emph {et~al.}(2015)\citenamefont {Frisch},
  \citenamefont {Mark}, \citenamefont {Aikawa}, \citenamefont {Baier},
  \citenamefont {Grimm}, \citenamefont {Petrov}, \citenamefont {Kotochigova},
  \citenamefont {Qu\'em\'ener}, \citenamefont {Lepers}, \citenamefont
  {Dulieu},\ and\ \citenamefont {Ferlaino}}]{PhysRevLett.115.203201}%
  \BibitemOpen
  \bibfield  {author} {\bibinfo {author} {\bibfnamefont {A.}~\bibnamefont
  {Frisch}}, \bibinfo {author} {\bibfnamefont {M.}~\bibnamefont {Mark}},
  \bibinfo {author} {\bibfnamefont {K.}~\bibnamefont {Aikawa}}, \bibinfo
  {author} {\bibfnamefont {S.}~\bibnamefont {Baier}}, \bibinfo {author}
  {\bibfnamefont {R.}~\bibnamefont {Grimm}}, \bibinfo {author} {\bibfnamefont
  {A.}~\bibnamefont {Petrov}}, \bibinfo {author} {\bibfnamefont
  {S.}~\bibnamefont {Kotochigova}}, \bibinfo {author} {\bibfnamefont
  {G.}~\bibnamefont {Qu\'em\'ener}}, \bibinfo {author} {\bibfnamefont
  {M.}~\bibnamefont {Lepers}}, \bibinfo {author} {\bibfnamefont
  {O.}~\bibnamefont {Dulieu}}, \ and\ \bibinfo {author} {\bibfnamefont
  {F.}~\bibnamefont {Ferlaino}},\ }\bibfield  {title} {\enquote {\bibinfo
  {title} {Ultracold dipolar molecules composed of strongly magnetic atoms},}\
  }\href {\doibase 10.1103/PhysRevLett.115.203201} {\bibfield  {journal}
  {\bibinfo  {journal} {Phys. Rev. Lett.}\ }\textbf {\bibinfo {volume} {115}},\
  \bibinfo {pages} {203201} (\bibinfo {year} {2015})}\BibitemShut {NoStop}%
\bibitem [{\citenamefont {Will}\ \emph {et~al.}(2016)\citenamefont {Will},
  \citenamefont {Park}, \citenamefont {Yan}, \citenamefont {Loh},\ and\
  \citenamefont {Zwierlein}}]{MartinNaK}%
  \BibitemOpen
  \bibfield  {author} {\bibinfo {author} {\bibfnamefont {Sebastian~A.}\
  \bibnamefont {Will}}, \bibinfo {author} {\bibfnamefont {Jee~Woo}\
  \bibnamefont {Park}}, \bibinfo {author} {\bibfnamefont {Zoe~Z.}\ \bibnamefont
  {Yan}}, \bibinfo {author} {\bibfnamefont {Huanqian}\ \bibnamefont {Loh}}, \
  and\ \bibinfo {author} {\bibfnamefont {Martin~W.}\ \bibnamefont
  {Zwierlein}},\ }\bibfield  {title} {\enquote {\bibinfo {title} {Coherent
  microwave control of ultracold $^{23}\mathrm{Na}^{40}\mathrm{K}$
  molecules},}\ }\href {\doibase 10.1103/PhysRevLett.116.225306} {\bibfield
  {journal} {\bibinfo  {journal} {Phys. Rev. Lett.}\ }\textbf {\bibinfo
  {volume} {116}},\ \bibinfo {pages} {225306} (\bibinfo {year}
  {2016})}\BibitemShut {NoStop}%
\bibitem [{\citenamefont {Nascimbene}\ \emph {et~al.}(2015)\citenamefont
  {Nascimbene}, \citenamefont {Goldman}, \citenamefont {Cooper},\ and\
  \citenamefont {Dalibard}}]{PhysRevLett.115.140401}%
  \BibitemOpen
  \bibfield  {author} {\bibinfo {author} {\bibfnamefont {Sylvain}\ \bibnamefont
  {Nascimbene}}, \bibinfo {author} {\bibfnamefont {Nathan}\ \bibnamefont
  {Goldman}}, \bibinfo {author} {\bibfnamefont {Nigel~R.}\ \bibnamefont
  {Cooper}}, \ and\ \bibinfo {author} {\bibfnamefont {Jean}\ \bibnamefont
  {Dalibard}},\ }\bibfield  {title} {\enquote {\bibinfo {title} {Dynamic
  optical lattices of subwavelength spacing for ultracold atoms},}\ }\href
  {\doibase 10.1103/PhysRevLett.115.140401} {\bibfield  {journal} {\bibinfo
  {journal} {Phys. Rev. Lett.}\ }\textbf {\bibinfo {volume} {115}},\ \bibinfo
  {pages} {140401} (\bibinfo {year} {2015})}\BibitemShut {NoStop}%
\bibitem [{\citenamefont {Boronat}\ and\ \citenamefont
  {Casulleras}(1994)}]{PhysRevB.49.8920}%
  \BibitemOpen
  \bibfield  {author} {\bibinfo {author} {\bibfnamefont {J.}~\bibnamefont
  {Boronat}}\ and\ \bibinfo {author} {\bibfnamefont {J.}~\bibnamefont
  {Casulleras}},\ }\bibfield  {title} {\enquote {\bibinfo {title} {{Monte
  Carlo} analysis of an interatomic potential for {He}},}\ }\href {\doibase
  10.1103/PhysRevB.49.8920} {\bibfield  {journal} {\bibinfo  {journal} {Phys.
  Rev. B}\ }\textbf {\bibinfo {volume} {49}},\ \bibinfo {pages} {8920--8930}
  (\bibinfo {year} {1994})}\BibitemShut {NoStop}%
\bibitem [{\citenamefont {Liu}\ \emph {et~al.}(1974)\citenamefont {Liu},
  \citenamefont {Kalos},\ and\ \citenamefont {Chester}}]{PhysRevA.10.303}%
  \BibitemOpen
  \bibfield  {author} {\bibinfo {author} {\bibfnamefont {K.~S.}\ \bibnamefont
  {Liu}}, \bibinfo {author} {\bibfnamefont {M.~H.}\ \bibnamefont {Kalos}}, \
  and\ \bibinfo {author} {\bibfnamefont {G.~V.}\ \bibnamefont {Chester}},\
  }\bibfield  {title} {\enquote {\bibinfo {title} {Quantum hard spheres in a
  channel},}\ }\href {\doibase 10.1103/PhysRevA.10.303} {\bibfield  {journal}
  {\bibinfo  {journal} {Phys. Rev. A}\ }\textbf {\bibinfo {volume} {10}},\
  \bibinfo {pages} {303--308} (\bibinfo {year} {1974})}\BibitemShut {NoStop}%
\bibitem [{\citenamefont {Casulleras}\ and\ \citenamefont
  {Boronat}(1995)}]{PhysRevB.52.3654}%
  \BibitemOpen
  \bibfield  {author} {\bibinfo {author} {\bibfnamefont {J.}~\bibnamefont
  {Casulleras}}\ and\ \bibinfo {author} {\bibfnamefont {J.}~\bibnamefont
  {Boronat}},\ }\bibfield  {title} {\enquote {\bibinfo {title} {Unbiased
  estimators in quantum {Monte Carlo} methods: Application to liquid
  $^{4}\mathrm{He}$},}\ }\href {\doibase 10.1103/PhysRevB.52.3654} {\bibfield
  {journal} {\bibinfo  {journal} {Phys. Rev. B}\ }\textbf {\bibinfo {volume}
  {52}},\ \bibinfo {pages} {3654--3661} (\bibinfo {year} {1995})}\BibitemShut
  {NoStop}%
\bibitem [{\citenamefont {Astrakharchik}\ \emph {et~al.}(2007)\citenamefont
  {Astrakharchik}, \citenamefont {Boronat}, \citenamefont {Casulleras},
  \citenamefont {Kurbakov},\ and\ \citenamefont
  {Lozovik}}]{PhysRevA.75.063630}%
  \BibitemOpen
  \bibfield  {author} {\bibinfo {author} {\bibfnamefont {G.~E.}\ \bibnamefont
  {Astrakharchik}}, \bibinfo {author} {\bibfnamefont {J.}~\bibnamefont
  {Boronat}}, \bibinfo {author} {\bibfnamefont {J.}~\bibnamefont {Casulleras}},
  \bibinfo {author} {\bibfnamefont {I.~L.}\ \bibnamefont {Kurbakov}}, \ and\
  \bibinfo {author} {\bibfnamefont {Yu.~E.}\ \bibnamefont {Lozovik}},\
  }\bibfield  {title} {\enquote {\bibinfo {title} {Weakly interacting
  two-dimensional system of dipoles: Limitations of the mean-field theory},}\
  }\href {\doibase 10.1103/PhysRevA.75.063630} {\bibfield  {journal} {\bibinfo
  {journal} {Phys. Rev. A}\ }\textbf {\bibinfo {volume} {75}},\ \bibinfo
  {pages} {063630} (\bibinfo {year} {2007})}\BibitemShut {NoStop}%
\bibitem [{\citenamefont {Bienaim\'e}\ \emph {et~al.}(2016)\citenamefont
  {Bienaim\'e}, \citenamefont {Fava}, \citenamefont {Colzi}, \citenamefont
  {Mordini}, \citenamefont {Serafini}, \citenamefont {Qu}, \citenamefont
  {Stringari}, \citenamefont {Lamporesi},\ and\ \citenamefont
  {Ferrari}}]{PhysRevA.94.063652}%
  \BibitemOpen
  \bibfield  {author} {\bibinfo {author} {\bibfnamefont {Tom}\ \bibnamefont
  {Bienaim\'e}}, \bibinfo {author} {\bibfnamefont {Eleonora}\ \bibnamefont
  {Fava}}, \bibinfo {author} {\bibfnamefont {Giacomo}\ \bibnamefont {Colzi}},
  \bibinfo {author} {\bibfnamefont {Carmelo}\ \bibnamefont {Mordini}}, \bibinfo
  {author} {\bibfnamefont {Simone}\ \bibnamefont {Serafini}}, \bibinfo {author}
  {\bibfnamefont {Chunlei}\ \bibnamefont {Qu}}, \bibinfo {author}
  {\bibfnamefont {Sandro}\ \bibnamefont {Stringari}}, \bibinfo {author}
  {\bibfnamefont {Giacomo}\ \bibnamefont {Lamporesi}}, \ and\ \bibinfo {author}
  {\bibfnamefont {Gabriele}\ \bibnamefont {Ferrari}},\ }\bibfield  {title}
  {\enquote {\bibinfo {title} {Spin-dipole oscillation and polarizability of a
  binary {Bose-Einstein} condensate near the miscible-immiscible phase
  transition},}\ }\href {\doibase 10.1103/PhysRevA.94.063652} {\bibfield
  {journal} {\bibinfo  {journal} {Phys. Rev. A}\ }\textbf {\bibinfo {volume}
  {94}},\ \bibinfo {pages} {063652} (\bibinfo {year} {2016})}\BibitemShut
  {NoStop}%
\bibitem [{\citenamefont {Kim}\ \emph {et~al.}(2017)\citenamefont {Kim},
  \citenamefont {Seo},\ and\ \citenamefont {Shin}}]{PhysRevLett.119.185302}%
  \BibitemOpen
  \bibfield  {author} {\bibinfo {author} {\bibfnamefont {Joon~Hyun}\
  \bibnamefont {Kim}}, \bibinfo {author} {\bibfnamefont {Sang~Won}\
  \bibnamefont {Seo}}, \ and\ \bibinfo {author} {\bibfnamefont
  {Y.}~\bibnamefont {Shin}},\ }\bibfield  {title} {\enquote {\bibinfo {title}
  {Critical spin superflow in a spinor {Bose-Einstein} condensate},}\ }\href
  {\doibase 10.1103/PhysRevLett.119.185302} {\bibfield  {journal} {\bibinfo
  {journal} {Phys. Rev. Lett.}\ }\textbf {\bibinfo {volume} {119}},\ \bibinfo
  {pages} {185302} (\bibinfo {year} {2017})}\BibitemShut {NoStop}%
\bibitem [{\citenamefont {Delehaye}\ \emph {et~al.}(2015)\citenamefont
  {Delehaye}, \citenamefont {Laurent}, \citenamefont {Ferrier-Barbut},
  \citenamefont {Jin}, \citenamefont {Chevy},\ and\ \citenamefont
  {Salomon}}]{PhysRevLett.115.265303}%
  \BibitemOpen
  \bibfield  {author} {\bibinfo {author} {\bibfnamefont {Marion}\ \bibnamefont
  {Delehaye}}, \bibinfo {author} {\bibfnamefont {S\'ebastien}\ \bibnamefont
  {Laurent}}, \bibinfo {author} {\bibfnamefont {Igor}\ \bibnamefont
  {Ferrier-Barbut}}, \bibinfo {author} {\bibfnamefont {Shuwei}\ \bibnamefont
  {Jin}}, \bibinfo {author} {\bibfnamefont {Fr\'ed\'eric}\ \bibnamefont
  {Chevy}}, \ and\ \bibinfo {author} {\bibfnamefont {Christophe}\ \bibnamefont
  {Salomon}},\ }\bibfield  {title} {\enquote {\bibinfo {title} {Critical
  velocity and dissipation of an ultracold {Bose-Fermi} counterflow},}\ }\href
  {\doibase 10.1103/PhysRevLett.115.265303} {\bibfield  {journal} {\bibinfo
  {journal} {Phys. Rev. Lett.}\ }\textbf {\bibinfo {volume} {115}},\ \bibinfo
  {pages} {265303} (\bibinfo {year} {2015})}\BibitemShut {NoStop}%
\bibitem [{\citenamefont {Castin}\ \emph {et~al.}(2015)\citenamefont {Castin},
  \citenamefont {Ferrier-Barbut},\ and\ \citenamefont
  {Salomon}}]{ComptesRendusPhysique.16.241}%
  \BibitemOpen
  \bibfield  {author} {\bibinfo {author} {\bibfnamefont {Yvan}\ \bibnamefont
  {Castin}}, \bibinfo {author} {\bibfnamefont {Igor}\ \bibnamefont
  {Ferrier-Barbut}}, \ and\ \bibinfo {author} {\bibfnamefont {Christophe}\
  \bibnamefont {Salomon}},\ }\bibfield  {title} {\enquote {\bibinfo {title}
  {The {Landau} critical velocity for a particle in a {Fermi} superfluid},}\
  }\href {\doibase 10.1016/j.crhy.2015.01.005} {\bibfield  {journal} {\bibinfo
  {journal} {Comptes Rendus Physique}\ }\textbf {\bibinfo {volume} {16}},\
  \bibinfo {pages} {241--253} (\bibinfo {year} {2015})}\BibitemShut {NoStop}%
\bibitem [{\citenamefont {Abad}\ \emph {et~al.}(2015)\citenamefont {Abad},
  \citenamefont {Recati}, \citenamefont {Stringari},\ and\ \citenamefont
  {Chevy}}]{EurPhysJD.69.126}%
  \BibitemOpen
  \bibfield  {author} {\bibinfo {author} {\bibfnamefont {Marta}\ \bibnamefont
  {Abad}}, \bibinfo {author} {\bibfnamefont {Alessio}\ \bibnamefont {Recati}},
  \bibinfo {author} {\bibfnamefont {Sandro}\ \bibnamefont {Stringari}}, \ and\
  \bibinfo {author} {\bibfnamefont {Fr{\'e}d{\'e}ric}\ \bibnamefont {Chevy}},\
  }\bibfield  {title} {\enquote {\bibinfo {title} {Counter-flow instability of
  a quantum mixture of two superfluids},}\ }\href {\doibase
  10.1140/epjd/e2015-50851-y} {\bibfield  {journal} {\bibinfo  {journal} {Eur.
  Phys. J. D}\ }\textbf {\bibinfo {volume} {69}},\ \bibinfo {pages} {126}
  (\bibinfo {year} {2015})}\BibitemShut {NoStop}%
\bibitem [{\citenamefont {Kravchenko}\ and\ \citenamefont
  {Fil}(2008)}]{JLowTempPhys.150.612}%
  \BibitemOpen
  \bibfield  {author} {\bibinfo {author} {\bibfnamefont {L.~Yu.}\ \bibnamefont
  {Kravchenko}}\ and\ \bibinfo {author} {\bibfnamefont {D.~V.}\ \bibnamefont
  {Fil}},\ }\bibfield  {title} {\enquote {\bibinfo {title} {Critical velocities
  in two-component superfluid {Bose} gases},}\ }\href {\doibase
  10.1007/s10909-007-9595-3} {\bibfield  {journal} {\bibinfo  {journal} {J. Low
  Temp. Phys.}\ }\textbf {\bibinfo {volume} {150}},\ \bibinfo {pages}
  {612--617} (\bibinfo {year} {2008})}\BibitemShut {NoStop}%
\bibitem [{\citenamefont {Kravchenko}\ and\ \citenamefont
  {Fil}(2009)}]{JLowTempPhys.155.219}%
  \BibitemOpen
  \bibfield  {author} {\bibinfo {author} {\bibfnamefont {L.~Y.}\ \bibnamefont
  {Kravchenko}}\ and\ \bibinfo {author} {\bibfnamefont {D.~V.}\ \bibnamefont
  {Fil}},\ }\bibfield  {title} {\enquote {\bibinfo {title} {Stationary waves in
  a supersonic flow of a two-component {Bose} gas},}\ }\href {\doibase
  10.1007/s10909-009-9879-x} {\bibfield  {journal} {\bibinfo  {journal} {J. Low
  Temp. Phys.}\ }\textbf {\bibinfo {volume} {155}},\ \bibinfo {pages}
  {219--234} (\bibinfo {year} {2009})}\BibitemShut {NoStop}%
\bibitem [{\citenamefont {Ishino}\ \emph {et~al.}(2011)\citenamefont {Ishino},
  \citenamefont {Tsubota},\ and\ \citenamefont
  {Takeuchi}}]{PhysRevA.83.063602}%
  \BibitemOpen
  \bibfield  {author} {\bibinfo {author} {\bibfnamefont {Shungo}\ \bibnamefont
  {Ishino}}, \bibinfo {author} {\bibfnamefont {Makoto}\ \bibnamefont
  {Tsubota}}, \ and\ \bibinfo {author} {\bibfnamefont {Hiromitsu}\ \bibnamefont
  {Takeuchi}},\ }\bibfield  {title} {\enquote {\bibinfo {title}
  {Countersuperflow instability in miscible two-component {Bose-Einstein}
  condensates},}\ }\href {\doibase 10.1103/PhysRevA.83.063602} {\bibfield
  {journal} {\bibinfo  {journal} {Phys. Rev. A}\ }\textbf {\bibinfo {volume}
  {83}},\ \bibinfo {pages} {063602} (\bibinfo {year} {2011})}\BibitemShut
  {NoStop}%
\bibitem [{\citenamefont {Hoefer}\ \emph {et~al.}(2011)\citenamefont {Hoefer},
  \citenamefont {Chang}, \citenamefont {Hamner},\ and\ \citenamefont
  {Engels}}]{PhysRevA.84.041605}%
  \BibitemOpen
  \bibfield  {author} {\bibinfo {author} {\bibfnamefont {M.~A.}\ \bibnamefont
  {Hoefer}}, \bibinfo {author} {\bibfnamefont {J.~J.}\ \bibnamefont {Chang}},
  \bibinfo {author} {\bibfnamefont {C.}~\bibnamefont {Hamner}}, \ and\ \bibinfo
  {author} {\bibfnamefont {P.}~\bibnamefont {Engels}},\ }\bibfield  {title}
  {\enquote {\bibinfo {title} {Dark-dark solitons and modulational instability
  in miscible two-component {Bose-Einstein} condensates},}\ }\href {\doibase
  10.1103/PhysRevA.84.041605} {\bibfield  {journal} {\bibinfo  {journal} {Phys.
  Rev. A}\ }\textbf {\bibinfo {volume} {84}},\ \bibinfo {pages} {041605}
  (\bibinfo {year} {2011})}\BibitemShut {NoStop}%
\bibitem [{\citenamefont {Hamner}\ \emph {et~al.}(2011)\citenamefont {Hamner},
  \citenamefont {Chang}, \citenamefont {Engels},\ and\ \citenamefont
  {Hoefer}}]{PhysRevLett.106.065302}%
  \BibitemOpen
  \bibfield  {author} {\bibinfo {author} {\bibfnamefont {C.}~\bibnamefont
  {Hamner}}, \bibinfo {author} {\bibfnamefont {J.~J.}\ \bibnamefont {Chang}},
  \bibinfo {author} {\bibfnamefont {P.}~\bibnamefont {Engels}}, \ and\ \bibinfo
  {author} {\bibfnamefont {M.~A.}\ \bibnamefont {Hoefer}},\ }\bibfield  {title}
  {\enquote {\bibinfo {title} {Generation of dark-bright soliton trains in
  superfluid-superfluid counterflow},}\ }\href {\doibase
  10.1103/PhysRevLett.106.065302} {\bibfield  {journal} {\bibinfo  {journal}
  {Phys. Rev. Lett.}\ }\textbf {\bibinfo {volume} {106}},\ \bibinfo {pages}
  {065302} (\bibinfo {year} {2011})}\BibitemShut {NoStop}%
\bibitem [{\citenamefont {Fava}\ \emph {et~al.}(2017)\citenamefont {Fava},
  \citenamefont {Bienaim\'e}, \citenamefont {Mordini}, \citenamefont {Colzi},
  \citenamefont {Qu}, \citenamefont {Stringari}, \citenamefont {Lamporesi},\
  and\ \citenamefont {Ferrari}}]{1708.03923}%
  \BibitemOpen
  \bibfield  {author} {\bibinfo {author} {\bibfnamefont {E.}~\bibnamefont
  {Fava}}, \bibinfo {author} {\bibfnamefont {T.}~\bibnamefont {Bienaim\'e}},
  \bibinfo {author} {\bibfnamefont {C.}~\bibnamefont {Mordini}}, \bibinfo
  {author} {\bibfnamefont {G.}~\bibnamefont {Colzi}}, \bibinfo {author}
  {\bibfnamefont {C.}~\bibnamefont {Qu}}, \bibinfo {author} {\bibfnamefont
  {S.}~\bibnamefont {Stringari}}, \bibinfo {author} {\bibfnamefont
  {G.}~\bibnamefont {Lamporesi}}, \ and\ \bibinfo {author} {\bibfnamefont
  {G.}~\bibnamefont {Ferrari}},\ }\bibfield  {title} {\enquote {\bibinfo
  {title} {Spin superfluidity of a {Bose} gas mixture at finite temperature},}\
  }\href@noop {} {\  (\bibinfo {year} {2017})},\ \Eprint
  {http://arxiv.org/abs/1708.03923} {arXiv:1708.03923} \BibitemShut {NoStop}%
\bibitem [{\citenamefont {Pitaevskii}\ and\ \citenamefont
  {Stringari}(2016)}]{book_pitaevskii.stringari.2016}%
  \BibitemOpen
  \bibfield  {author} {\bibinfo {author} {\bibfnamefont {Lev}\ \bibnamefont
  {Pitaevskii}}\ and\ \bibinfo {author} {\bibfnamefont {Sandro}\ \bibnamefont
  {Stringari}},\ }\href@noop {} {\emph {\bibinfo {title} {Bose-Einstein
  condensation and superfluidity}}}\ (\bibinfo  {publisher} {Oxford University
  Press},\ \bibinfo {year} {2016})\BibitemShut {NoStop}%
\end{thebibliography}%

\appendix

\section{Quantum hydrodynamics for a non-symmetric mixture}
\label{app:nonZ2}

We consider Eq.~\eqref{eq:Hhydro} in the more general case in which the two 
species are not symmetric ($\rho_{11} \neq \rho_{22}$, $\alpha_{11} \neq 
\alpha_{22}$ and $m_1 \neq m_2$).
Hamilton equations lead to the equation of motion for phase fluctuations,
\begin{equation}
 \ddot{\phi_i}(x) = - \left( \frac{\hbar^2}{m_i m_l} \alpha_{il} \rho_{lj} 
\right) k^2 \phi_j(x),
\end{equation}
and the quantity in parentheses can be identified with the matrix $C_{ij}$ of 
the squared speeds of sound.

Written with respect to an eigenbasis of $C_{ij}$, the system exhibits two 
non-interacting modes which obey a linear dispersion relation $\omega_\pm = 
c^2_\pm k^2$. The speeds of sound $c_\pm^2$, eigenvalues of $C_{ij}$,   
generalise Eqs.~\eqref{eq:cd}-\eqref{eq:cs}. Obviously, contrary to Sec.\ 
\ref{sec:quantum_hydro}, the two independent modes are not ``pure" 
density and spin channels, due to the lack of $\mathbb{Z}_2$ symmetry.

In terms of the coefficients of matrix $C_{ij}$, the eigenspeeds of sound read
\begin{align}
 c^2_\pm = & \frac{C_{11} + C_{22}}{2} \nonumber \\
 & \pm \frac{1}{2}
 \sqrt{(C_{11} + C_{22})^2 - 4 (C_{11} C_{22} - C_{12} C_{21})},
\end{align}
The system is stable as long as $c_\pm^2$ are positive, which leads to the 
inequality 
$C_{11}C_{22} - C_{12}C_{21} >0$, or, in terms of the model's coefficients,
\begin{equation}
 (\alpha_{11} \alpha_{22} - \alpha_{12}^2)
 (\rho_{11} \rho_{22} - \rhodrag^2) > 0
\end{equation}
The second factor in this equation is always positive, thanks to the bound on 
the magnitude of the superfluid drag (cf.\ discussion leading to
Eq.~\eqref{eq:drag_bound}). We thus obtain a condition on the strength of the 
density-density interactions, namely
$(\alpha_{11} \alpha_{22} - \alpha_{12}^2)>0$, thus recovering the well known
criterion for the onset of the phase separation instability 
\cite{book_pitaevskii.stringari.2016}.

\section{Relation between winding numbers and superfluid densities}
\label{sec:app_path_integral}

Following \cite{PhysRevB.36.8343}, we imagine a slab of fluid sandwiched
between two parallel boundaries, moving with respect to the fluid with velocity
$\vec{v}$. We can write the density matrix of a two-species system in a frame
comoving with the boundaries as
\begin{align}
&\rho_v(R, R', \beta) = \bra{R'} e^{-\beta H_v} \ket{R}, \\
&H_v = \sum_{\alpha=\{1,2\}}\sum_{j=1}^{N_\alpha} \frac{(\vec{p}_j -
m_\alpha \vec{v})^2}{2m_\alpha} + V,
\end{align}
where $V$ comprises both inter- and intra-species interactions, both assumed 
to depend on positions only. (Greek indices identify the species, while Latin 
indices count particles). The change in total momentum in the frame of 
reference of the boundaries is due to the response of the normal component 
$\rho_N$ of the fluid, hence, calling $\rho_0 = \rho_{v=0}$, the density 
matrix in the frame of rest of the fluid,
\begin{equation}
 \frac{\rho_N}{\rho_0} (N_1 m_1 + N_2 m_2)\vec{v} =
 \frac{\rho_N}{\rho_0} M \vec{v} = \avg{\vec{P}_v} =
 \frac{\tr [\vec{P}_v \rho_v]}{\tr[\rho_v]},
\end{equation}
and is equal to one minus the total superfluid fraction, which can then be
written in terms of the variation of the free energy as
\begin{equation} \label{eq:rhos_free_energy_shift}
 \frac{\rho_T}{\rho_0} = \frac{\partial
\mathcal{F}_v}{\partial(\frac{1}{2} M v^2)}.
\end{equation}

We can write the density matrix in Fourier space and relate the density matrix
$\rho_0$ in the frame of rest to $\rho_v$,
\begin{equation}
 \rho_v = \exp \left\lbrace i \vec{v} \cdot \sum_{\alpha, j}
(r_{j\alpha} - r'_{j\alpha}) m_\alpha\right\rbrace \tilde \rho_0.
\end{equation}
Under periodic permutations of the particles,
\begin{equation}
 \rho_v = e^{i \vec{v} \cdot m_1 \vec{W_1} L}
 e^{i \vec{v} \cdot m_2 \vec{W_2}L} \tilde \rho_0,
\end{equation}
whence
\begin{align}
 e^{-\beta \Delta \mathcal{F}_v} &= \frac{1}{\mathcal{Z}}
    \int \tilde \rho_0\, e^{i (m_1 \vec{W_1} + m_2 \vec{W_2})\cdot \vec{v}L}
\nonumber \\
& = \avg{e^{i (m_1 \vec{W_1} + m_2 \vec{W_2})\cdot \vec{v}L}}.
\end{align}
Expanding to leading order the previous expression, one finally gets
\begin{align}
 \Delta \mathcal{F}_v = \frac{1}{2} \frac{v^2 L^2}{d\beta}
    \avg{(m_1 W_1 + m_2 W_2)^2}
\end{align}
and, using Eq.~\eqref{eq:rhos_free_energy_shift} and $\rho_0 = M L^{-d}$,
\begin{align}
 \rho_s^{tot} & = \frac{L^{2-d}}{d\beta} \avg{(m_1 W_1 + m_2 W_2)^2},
\end{align}
yielding Eq.~\eqref{eq:winding_drag}.

\vfill

\end{document}